\documentclass[traditabstract]{aa}

\usepackage{natbib}
\usepackage[varg]{txfonts}
\usepackage{graphicx}
\usepackage{url}

\bibpunct{(}{)}{;}{a}{}{,}

\usepackage{color,soul}
\newcommand{\edit}[1]{#1}
\newcommand{\editm}[1]{#1}

\begin{document}

\title{Chemical modeling of the \object{L1498} and \object{L1517B}
  prestellar cores: \\ \edit{CO and HCO$^{+}$ depletion}\thanks{Based on
    observations carried out with the IRAM 30m Telescope. IRAM is
    supported by INSU/CNRS (France), MPG (Germany) and IGN (Spain).}}

\author{S. Maret \inst{1} \and E.~A. Bergin \inst{2} \and
  M. Tafalla \inst{3}}

\titlerunning{Chemical modeling of the L1498 and L1517B prestellar
  cores}

\institute{UJF-Grenoble 1 / CNRS-INSU, Institut de Plan\'etologie et
  d'Astrophysique de Grenoble (IPAG) UMR 5274, Grenoble, F-38041,
  France \and Department of Astronomy, University of Michigan, 500
  Church Street, Ann Arbor MI-48104, USA \and Observatorio Astronómico
  Nacional, Alfonso XII 3, 28014 Madrid, Spain}

\date{Received ...; accepted ...}
 
\abstract{Prestellar cores \edit{exhibit} a strong chemical
  differentiation, which is mainly caused by the freeze-out of
  molecules onto the grain surfaces. Understanding this chemical
  structure is important, because molecular lines are often used as
  probes to constrain the core physical properties. Here we present
  new observations and analysis of the C$^{18}$O~(1-0) and
  H$^{13}$CO$^{+}$~(1-0) line emission in the L1498 and L1517B
  prestellar cores, located in the Taurus-Auriga molecular complex. We
  model these observations with a detailed chemistry network coupled
  \edit{to a} radiative transfer code. Our model successfully
  reproduces the observed C$^{18}$O~(1-0) emission for a chemical age
  of a few 10$^{5}$~years. On the other hand, the observed
  H$^{13}$CO$^{+}$~(1-0) is reproduced only if cosmic-ray desorption
  by secondary photons is included, and if the grain\edit{s} have
  grown to a bigger size than average ISM grains in the core
  interior. This grain growth is consistent \edit{with} the
  \edit{infrared scattered light ("coreshine")} detected in these two
  objects, and is found to increase the CO abundance in the core
  interior by about a factor four. \edit{According to our model, CO is
    depleted by about 2-3 orders of magnitude in the core center.}}

\keywords{Astrochemistry -- ISM: abundances -- ISM: individual (L1498)
  -- ISM: individual (L1517B) -- ISM: molecules -- Stars: formation}
\maketitle

\section{Introduction}
\label{sec:introduction}

Prestellar cores represent the earliest stage \edit{in} the formation
of a star. The study of their physical properties, such as their
density, temperature and kinematics, is therefore important to
\edit{determine} how they form and evolve, and in turn to place
constraints on star formation theories. Molecular lines are often used
to derive core physical properties \citep[see
e.g.][]{Bergin07b}. However, because the core interior is dense
\citep[$\sim 10^{5}-10^{6}$~cm$^{-3}$, e.g. ][]{Alves01,Tafalla02} and
cold \citep[$\sim 5-10$~K; ][]{Bergin06b,Crapsi07}, most species
freeze-out on dust grains. This causes a strong chemical
differentiation, with the outer \edit{layers} of the cores being
relatively undepleted, and the abundances of heavy \edit{species} in
the \edit{central} regions being, on the other hand, reduced by up to
several orders of magnitude
\citep{Caselli99,Tafalla02,Tafalla04,Bergin02a}. Not all species are
sensitive to depletion: for example, some nitrogen-bearing species
such as N$_{2}$H$^{+}$ and NH$_{3}$ appear to \edit{remain} longer in
the gas phase \citep{Bergin02a,Maret06,Tafalla06}. A consequence of
the depletion of the heavy \edit{species} is \edit{an abundance
  increase} of deuterated species \citep{Bacmann03,Pagani07}. In order
to use molecular lines as ``probes'' of the physical conditions in the
core interiors, it is necessary to \edit{first} understand their
chemical structure. In addition, the amount of depletion should
increase with the core age. Therefore the chemical structure of a core
could in principle be used to constrain the core age \citep[see
e.g.][]{Bergin02a}.

L1498 and L1517B are two prestellar cores located in the Taurus-Auriga
complex. Because \edit{of} their close-to-round shapes, these two
cores have been studied extensively by \citet{Tafalla04,Tafalla06},
who mapped their dust continuum emission as well as several
lines. \citeauthor{Tafalla04} modeled the dust continuum emission
finding that their density distribution is consistent with that of
isothermal spheres at (or close to) equilibrium. In addition,
\edit{these authors} constrained the gas temperature from NH$_{3}$
rotation-inversion lines and non-thermal line widths from several
other lines, and concluded that thermal support dominates over
turbulent support. \citet{Kirk06} \edit{measured} the magnetic field
in these two cores from dust polarization measurements and found that
thermal support also dominates over the magnetic plus turbulent
support. In addition, \citet{Tafalla06} carried out a molecular survey
of L1498 and L1517B, suggesting that the majority of species (CO, CS,
CH$_3$OH, HCO$^{+}$, etc.) with the exception of NH$_3$ and
N$_{2}$H$^{+}$ are depleted in the core interiors. They modeled the
emission of these lines assuming simple abundance profiles (step
functions), and found that most species present sharp central holes.

In this paper, we present new observations of the C$^{18}$O~(1-0) and
H$^{13}$CO$^{+}$~(1-0) emission in these two cores. We attempt to
match these observations with a time-dependent chemical model
(including depletion and carbon fractionation) coupled to a non-LTE
radiative transfer code, with the aim of deriving the core chemical
structure and to place constraints on the core age.  The paper is
organized as follows. Observations are presented in
\S~\ref{sec:observations}. The model used to interpret the
observations is detailed in \S~\ref{sec:model}. Our results are
presented in \S~\ref{sec:results} and the implications of these
findings are discussed in \S~\ref{sec:discussion}. Finally
\S~\ref{sec:conclusions} concludes the article.

\begin{figure*}
  \centering
  \includegraphics[width=17.5cm]{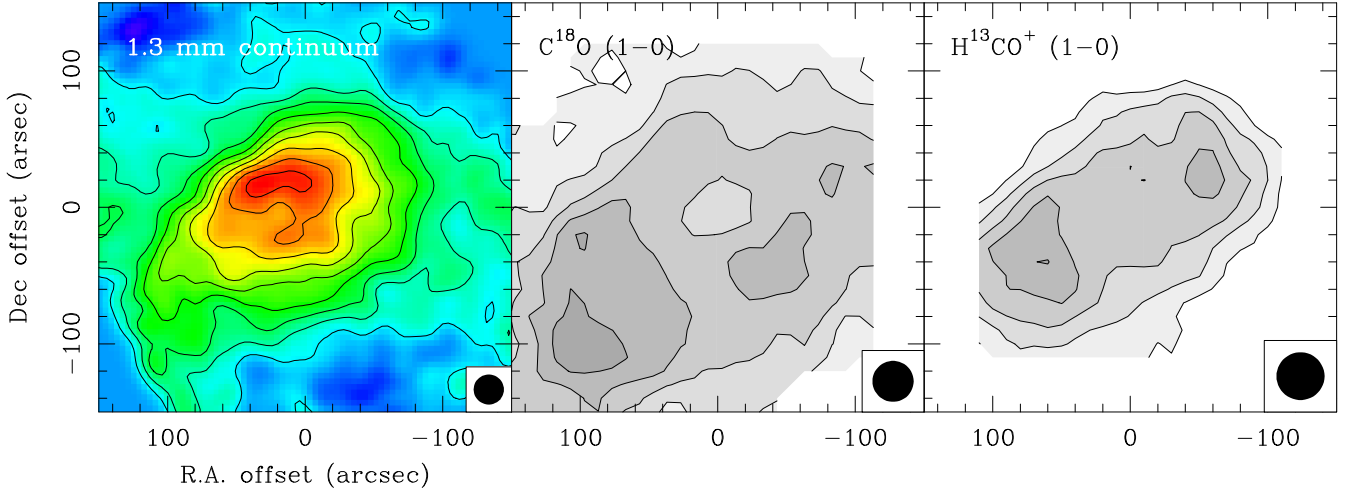}
  \caption{\emph{Left panel:} 1.3~mm continuum emission observed in
    L1498 \citep{Tafalla04}. Contour levels start at 2~mJy per
    \edit{11}\arcsec~beam and increase by steps of 2~mJy per
    \edit{11}\arcsec~beam. \emph{Middle panel:} C$^{18}$O~(1-0) integrated
    intensity. Contour levels starts at 0.2~K~km~s$^{-1}$
    \edit{(6.2$\sigma$)} and increase by steps of
    0.2~K~km~s$^{-1}$. \emph{Right panel:} H$^{13}$CO$^{+}$~(1-0)
    integrated intensity. Levels start at 0.1~K~km~s$^{-1}$
    \edit{(5.9$\sigma$)} and increase by steps of
    0.1~K~km~s$^{-1}$. In \edit{each} panel the spatial resolution of
    the map is indicated by \edit{a} filled circle.}
  \label{fig:l1498-map}
\end{figure*}

\section{Observations and data reduction}
\label{sec:observations}

We observed L1498 and L1517B in June 2012 with the IRAM 30m telescope
\edit{under average summer weather conditions}. The C$^{18}$O~(1-0)
and H$^{13}$CO$^{+}$ (1-0) line emission (109.78 and 86.75 GHz,
respectively) were mapped over a region $4' \times 4'$ in each core
(except for the C$^{18}$O~(1-0) map in L1498 which is $5' \times 5'$)
centered on $\alpha=4^{h}10^{m}51.5^{s}$ and
$\delta=25\degr09\arcmin58\arcsec$ for L1498, and on
$\alpha=4^{h}55^{m}18.8^{s}$ and $\delta=30\degr38\arcmin04\arcsec$
(J2000) for L1517B. These maps were obtained in on-the-fly mode with
position-switching, with a reference position offset with respect to
the map center of (-10\arcmin;-10\arcmin) for L1498, and
(0\arcmin;+10\arcmin) in L1517B. We checked that the reference
positions were free of C$^{18}$O~(1-0) and H$^{13}$CO$^{+}$ (1-0)
emission using frequency switching mode observations. The EMIR
receivers were used together with the VESPA autocorrelator. A
\edit{separate} receiver setting was used for the two
lines. \edit{Typical system temperatures were 100~K and 150~K for the
  H$^{13}$CO$^{+}$ (1-0) and C$^{18}$O~(1-0) receiver settings,
  respectively.} The autocorrelator was set to a spectral resolution
of 6.6~kHz for both lines, i.e. about $\sim~0.02$~km~s$^{-1}$ in
velocity. Observations were corrected for atmospheric transmission
using the standard chopper wheel technique. Pointing was checked
regularly on Uranus and strong quasars, and \edit{its RMS} accuracy
was found to be 2\arcsec .  The beam size of the IRAM 30m telescope is
28\arcsec\ and 22\arcsec\ for H$^{13}$CO$^{+}$ (1-0) and
C$^{18}$O~(1-0), respectively.

Data reduction was done using the CLASS package, part of the GILDAS
software suite\footnote{\url{http://www.iram.fr/IRAMFR/GILDAS}}. After
removing a polynomial baseline on each spectrum, the data were gridded
to produce spectral data cubes. In order to improve the
signal-to-noise ratio, the data were re-sampled at a spatial
resolution of 35\arcsec\ and 30\arcsec\ for H$^{13}$CO$^{+}$ (1-0) and
C$^{18}$O~(1-0), respectively. In addition, the data were re-sampled
at a spectral resolution of $\sim 0.05$~km~s$^{-1}$ for both lines.
\edit{The noise per channel on the final data cubes is 80~mK for
  H$^{13}$CO$^{+}$~(1-0), and 140~mK for C$^{18}$O~(1-0)}.  The data
were then converted to main beam temperature scale assuming a beam
efficiency of 0.81 and a forward efficiency of 0.95 for both lines, as
recommended on the IRAM website. Finally, the velocity channels were
summed in order to produce velocity integrated maps.

\edit{Figure}~\ref{fig:l1498-map} shows \edit{the} C$^{18}$O~(1-0) and
H$^{13}$CO$^{+}$~(1-0) line maps we obtained \edit{for} L1498,
together with the 1.3~mm continuum from \cite{Tafalla04}.  In the
continuum map, the core has an ellipsoidal shape with \edit{the} major
axis oriented along the Southeast-Northwest direction. The
C$^{18}$O~(1-0) emission appears anti-correlated to the dust continuum
emission, \edit{forming} a ring-like structure \edit{that surrounds}
the dust peak. This is a clear indication of gas freeze-out at the
center of the core. As already mentioned, such depletion ``holes''
have been observed in several other molecules towards L1498 and L1517B
\citep{Willacy98,Tafalla04,Tafalla06}, \edit{and} also towards other
cores such as \object{B68} or \object{L1544}
\citep{Caselli99,Bergin02a,Tafalla02}. The H$^{13}$CO$^{+}$~(1-0)
emission is somewhat similar to that of C$^{18}$O~(1-0): the emission
is not centered on the dust continuum peak, but instead has two maxima
to the Northwest and the Southeast of the core center. This is likely
a result of the depletion of CO and its isotopologues, because
$^{13}$CO is a parent molecule of H$^{13}$CO$^{+}$~(1-0).

\begin{figure*}
  \centering
  \includegraphics[width=17.5cm]{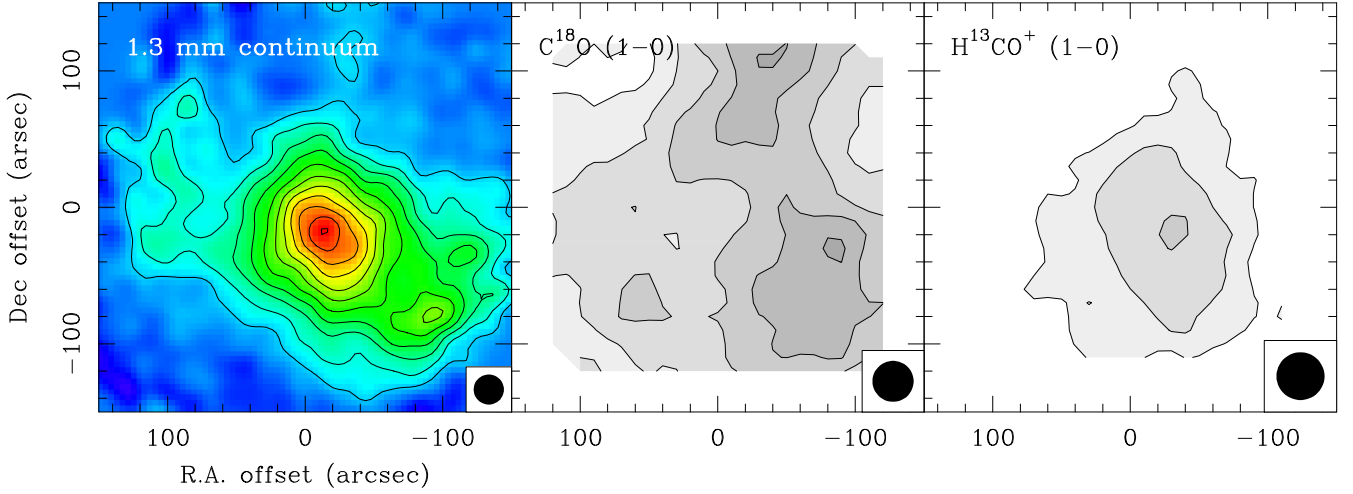}
  \caption{\edit{\emph{Left panel:} 1.3 mm continuum emission observed
      in L1517B. \emph{Middle panel:} C$^{18}$O~(1-0) integrated
      intensity. \emph{Right panel: } H$^{13}$CO$^{+}$~(1-0)
      integrated intensity. Contour levels are the same as in
      Fig.~\ref{fig:l1498-map}.}}
  \label{fig:l1517b-map}
\end{figure*}

\edit{Figure}~\ref{fig:l1517b-map} presents the same comparison for
the L1517B core. In this core the dust emission is more compact and
roundish than in L1498. Like in L1498, the C$^{18}$O~(1-0) emission
is, to some extent, anti correlated with the dust emission: the
position of the dust maximum emission corresponds to a dip in the
C$^{18}$O~(1-0) emission. However, the C$^{18}$O~(1-0) emission is
much more asymmetric than in L1498. In this case, most of the
C$^{18}$O~(1-0) emission is located to the \edit{West} of the core,
with little emission \edit{towards} the \edit{East} side. The
H$^{13}$CO$^{+}$~(1-0) emission stands in sharp contrast with the
C$^{18}$O~(1-0) distribution: it has an elongated shape roughly
centered on the dust continuum peak. Unlike in L1498, the
H$^{13}$CO$^{+}$~(1-0) does not follow the C$^{18}$O~(1-0) emission,
but appear\edit{s} somewhat anti-correlated, and closer to the dust
distribution.  In the following, we model C$^{18}$O~(1-0) and
H$^{13}$CO$^{+}$~(1-0) line emission\edit{s} in the two cores with a
chemistry model coupled with a radiative transfer code.

\section{Model}
\label{sec:model}

\subsection{Physical model}
\label{sec:physical-model}

We use the density profiles derived for the two cores by
\cite{Tafalla04}, \edit{who} modeled radial profile of the continuum
emission and found that these are well reproduced by density profiles
of the form:

\begin{equation}
  n(r) = \frac{n_{0}}{1 + (r / r_0)^\alpha}
\end{equation}

\noindent
where $n(r)$ is the H$_{2}$ density, $r$ the core radius, $n_{0}$ the
central density, and $r_{0}$ the \edit{half maximum density
  radius}. The values of $n_{0}$, $r_{0}$ and $\alpha$ they obtained
for the two cores are given in
Table~\ref{tab:density-profiles}. \edit{We note that these density
  profiles depend on the assumed dust emissivity, which is rather
  uncertain and may vary with the density. In addition, they depend on
  the assumed geometry for the cores: although the two cores appears
  ``roundish'', the line of sight dimension of the cores may be
  different than that observed in the plane of the sky. However,
  models using these the density profiles fit a number of line
  observations with different critical densities, two of them being
  the $\mathrm{\editm{N_{2}H^{+}~(3-2)}}$ and
  $\mathrm{\editm{N_{2}H^{+}~(1-0)}}$ lines towards the core center
  \citep[see][Fig.~6 and 7]{Tafalla04}. In addition, \cite{Keto04}
  modeled the $\mathrm{\editm{N_{2}H^{+}~(3-2)}}$ and
  $\mathrm{\editm{N_{2}H^{+}~(1-0)}}$ spectra in L1517B with a
  detailed radiative transfer model that includes the hyperfine
  components of this species, and they obtained a core central density
  of $1.3 \times 10^{5}$~cm$^{-3}$, in good agreement with the value
  obtained \citeauthor{Tafalla04}. To the best of our knowledge, the
  density profiles for the two cores are consistent with all published
  observations.} Because the two cores are embedded in an extended
molecular cloud, the density profiles are truncated at $r_{1} = 2.8
\times 10^{4}$~AU (200\arcsec\ at 140~pc) and we assume that the
density remains constant between $r_{1}$ and the external radii
$r_{2}$; the density in the region between $r_{1}$ and $r_{2}$
(hereafter the halo) is noted $n_{1}$. The values of $n_{1}$ and
$r_{2}$ are adjusted to reproduce the observed extended
C$^{18}$O~(1-0) emission, as explained later. The cores are assumed to
be isothermal with a temperature of 10~K. Furthermore the gas and the
dust grains are assumed to be thermally coupled. Finally we adopt a
distance of 140~pc for both sources.

\begin{table}
  \caption{Density profile parameters}
  \label{tab:density-profiles}
  \centering
  \begin{tabular}{lll}
    \hline
    \hline
    & L1498 & L1517B\\
    \hline
    $n_{0}$ (cm$^{-3}$) & $9.4 \times 10^{4}$ & $2.2 \times 10^{5}$\\
    $r_{0}$ (AU) & $1.0 \times 10^{4}$ & $4.9 \times 10^{3}$ \\
    $\alpha$ & 3.5 & 2.5\\
    \hline
    $r_{1}$ (AU) & $2.8 \times 10^{4}$ & $2.8 \times 10^{4}$ \\ 
    $r_{2}$ (AU) & $1.0 \times 10^{5}$ & $\editm{5.9 \times 10^{4}}$ \\ 
    $n_{1}$ (cm$^{-3}$) & $2.0 \times 10^{3}$ & $2.0 \times 10^{3}$ \\
    %$A_{v}$ & 2.3 & 1.2\\
    \hline
  \end{tabular}
\end{table}

We assume the cores to be static and we do not consider any temporal
variation of the density and temperature
profile\edit{s}. \cite{Tafalla04} measured the non-thermal line width
in L1498 and L1517B from hyperfine fits of the NH$_{3}$ and
N$_{2}$H$^{+}$ lines, and they did not \edit{find} any systematic
variation as a function of offset from the core centers. Therefore we
assume a constant turbulent velocity with zero velocity gradient. This
turbulent velocity is adjusted in order to reproduce the observed line
width, as we will see below.

\subsection{Chemical model}
\label{sec:chemical-model}

The chemical abundances are followed as a function of time \edit{and}
radius in core.  For this, we use the {\tt astrochem} code (Maret \&
Bergin, in prep.). This code is a time dependent gas-phase chemistry
model that also includes gas grain interactions (such as freeze-out
and desorption). It is similar to the code of \citet{Bergin97} that we
used in our previous studies of B68
\citep{Bergin02a,Bergin06b,Maret06,Maret07a,Maret07b}, but it uses a
faster ordinary differential equation solver ({\tt SUNDIALS};
\citealt{Hindmarsh05}). The code was benchmarked against the {\tt
  ALCHEMIC} code \citep{Semenov10}, and both were found to be in
excellent agreement.

We use the 2008 version of the gas phase reaction network from the
Ohio State University. This network includes over 6000 reactions
involving 650 species. In order to model the H$^{13}$CO$^{+}$
emission, we consider the isotopic fraction of $^{13}$C through
ion-neutral reactions. At low temperature, carbon fractionation occurs
mostly through the following reactions \citep{Langer84}:

\begin{equation}
  \mathrm{^{13}C^{+} + ^{12}CO \rightleftharpoons ^{12}C^{+} +
    ^{13}CO + 35~K}
  \label{eq:3}
\end{equation}

\begin{equation}
  \mathrm{H^{12}CO^{+} + ^{13}CO \rightleftharpoons H^{13}CO^{+} + 
    ^{12}CO + 9~K}
  \label{eq:4}
\end{equation}

For these two reactions, we use the forward and backward reaction
rates at 10~K from \citet{Langer84}. For all other reactions involving
carbon bearing species, we assume the same rate than for the main
isotopologue reaction, multiplied by a statistical branching ratio if
two of more products of the reaction contains a carbon atom. Selective
CO photo-dissociation may also contribute to carbon fractionation
\citep{Lee08}. However, this process is inefficient at $A_{v} > 1$
from where most of the observed C$^{18}$O~(1-0) and
H$^{13}$CO$^{+}$~(1-0) arise, and we have therefore neglected
it. Oxygen fractionation is also neglected; we merely assume that the
$\mathrm{C^{16}O/C^{18}O}$ ratio is equal to the elemental
$\mathrm{^{16}O/^{18}O}$ ratio. \edit{This is consistent with detailed
  fractionation models for similar conditions \citep{Langer84}.}  In
our model, we assume that \citep[$\mathrm{^{12}C/^{13}C} = 70$ and
$\mathrm{^{16}O/^{18}O}=560$; ][]{Wilson94}.

As mentioned above, several gas grain interactions are taken into
account in our model. All neutral species in the network are allowed
to freeze-out on dust grains. The \edit{freeze-out} rate depends both
on the grain abundance and geometrical cross section. For our first
models, we assume spherical dust grains with a radius of 0.1~$\mu$m
and composed of olivine; the influence of the grain size is
investigated later in the paper. Assuming a dust to gas mass ratio of
100, this gives a grain number density of $1.32 \times 10^{-12}$ with
respect to H nuclei. Grains are assumed to be thermally coupled with
the gas. Both thermal, UV and cosmic-ray induced desorption are
considered. For thermal and cosmic-ray desorption, we adopt the
formalism developed by \citet{Hasegawa93a}. However, for H$_{2}$O and
CO cosmic-ray induced desorption, we \edit{adopt} the more recent
results \edit{by} \citet{Bringa04}. Thermal and cosmic-ray desorption
rates depend critically on the binding energy of species on the grain
mantles. For these, we use the values from \citet{Garrod06}. For UV
desorption, we adopt the formalism of \citet{Oberg09b,Oberg09a}. The
photodesorption rate depends on the desorption yield, which has been
measured in the laboratory only for a handful of species. For all
other species, we adopt a yield of 10$^{-3}$~molecule per incident UV
photon. In our first models, we neglect the desorption from secondary
UV photons, but their effect is also investigated later in the
paper. \edit{Finally, we also consider electron attachment on grains,
  as well as recombination of ions onto charged grains, following
  \citet{Semenov10}}.

The initial abundances are given in
Table~\ref{tab:initial-abundances}. These are essentially the same
than in our previous studies \citep[see ][]{Maret07a}. We assume that
carbon monoxide and water are pre-existing, with water in form of
ices. Nitrogen is assumed to be mostly in atomic form \citep{Maret06},
and iron is assumed to be heavily depleted \citep{Maret07a}. However,
we use a slightly different helium abundance, equal to that of the Sun
\citep{Asplund09}. In addition, we consider the sulfur and silicon
chemistry. For these two element\edit{s} we adopt the atomic
abundances measured in $\zeta$~Ophiuchi \citep{Savage96}, depleted by
two orders of magnitude. This large depletion factor is needed
\edit{to} reproduce the abundance of molecular ions \citep[see
e.g.][]{Graedel82}. S, Si and Fe, whose ionization potential is lower
than that of H (13.6~eV), are assumed to be ionized. Other species are
assumed to be neutral. Two important parameters of the models are the
core external UV radiation field and the cosmic-ray ionization rate;
these are left as free parameters and are adjusted from the
observations.

\begin{table}
  \caption{Initial abundances}
  \label{tab:initial-abundances}
  \centering
  \begin{tabular}{ll}
    \hline
    \hline
    Species & Abundance\tablefootmark{a}\\
    \hline
    H$_{2}$       & $0.5$\\
    He            & $8.5 \times 10^{-2}$\\
    H$_{2}$O ices & $2.2 \times 10^{-4}$\\
    CO            & $8.5 \times 10^{-5}$\\
    $^{13}$CO     & $\editm{1.2 \times 10^{-6}}$\\
    N             & $1.5 \times 10^{-5}$\\
    N$_{2}$       & $2.5 \times 10^{-6}$\\
    S$^{+}$       & $2.8 \times 10^{-7}$\\
    Si$^{+}$      & $1.7 \times 10^{-8}$\\
    Fe$^{+}$      & $3.0 \times 10^{-9}$\\
    $e^{-}$       & $\editm{3.0 \times 10^{-7}}$\\
    \hline
  \end{tabular}
  \tablefoot{
    \tablefoottext{a}{Abundances are given with respect to H nuclei.}
    }
\end{table}

\subsection{Radiative transfer model}
\label{sec:radiative-transfer}

In order to compute the line emission from the cores, we use the
non-LTE Monte-Carlo 1D (spherical) radiative transfer-code {\tt
  RATRAN} \citep{Hogerheijde00b}. The C$^{18}$O and H$^{13}$CO$^{+}$
abundances computed with our chemical model at each radius and time
step are used as an input of the radiative transfer code. We use the
Einstein coefficients and energy levels from the Cologne Molecular
Database for Spectroscopy \citep[CDMS; ][]{Muller01}. For both species
the same collisional rate\edit{s} as for the main isotopologue are
adopted. For C$^{18}$O, we use the CO collisional rates with
\edit{ortho and para} H$_{2}$ of \citet{Yang10}. Indirect measurements
of the H$_{2}$ ortho-to-para ratio in dense cores indicate ratios in
the range $10^{-2}-10^{-1}$ \citep{Maret07a,Pagani09}, and we
therefore assume that all H$_{2}$ is in para form. For
H$^{13}$CO$^{+}$ we use the HCO$^{+}$ collisional rates with
\edit{para} H$_{2}$ of \citet{Flower99b}. The H$^{13}$CO$^{+}$~(1-0)
line has an hyperfine structure that slightly broadens this line
\citep{Schmid-Burgk04}. To take this effect into account, we add a
0.133~km~s$^{-1}$ broadening in quadrature to the turbulent
broadening, following \citet{Tafalla06}. All molecular data files were
retrieved from the Leiden Atomic and Molecular Database \citep[LAMDA;
][]{Schoier05a}.

The radiative transfer model is used to produce synthetic data cubes
for both the C$^{18}$O~(1-0) and H$^{13}$CO$^{+}$~(1-0) lines. These
data cubes are then convolved with a 25\arcsec\ and 30\arcsec\ FWHM
Gaussian respectively, i.e. the effective (after spatial smoothing,
see \S~\ref{sec:observations}) resolution of our
observations. Finally, the convolved data cubes are compared to the
observations.

\section{Results}
\label{sec:results}

\subsection{$\mathrm{C^{18}O}$ emission}
\label{sec:c18o-emission}

\begin{figure*}
  \centering
  \begin{tabular}{cc}
    \includegraphics[width=8cm]{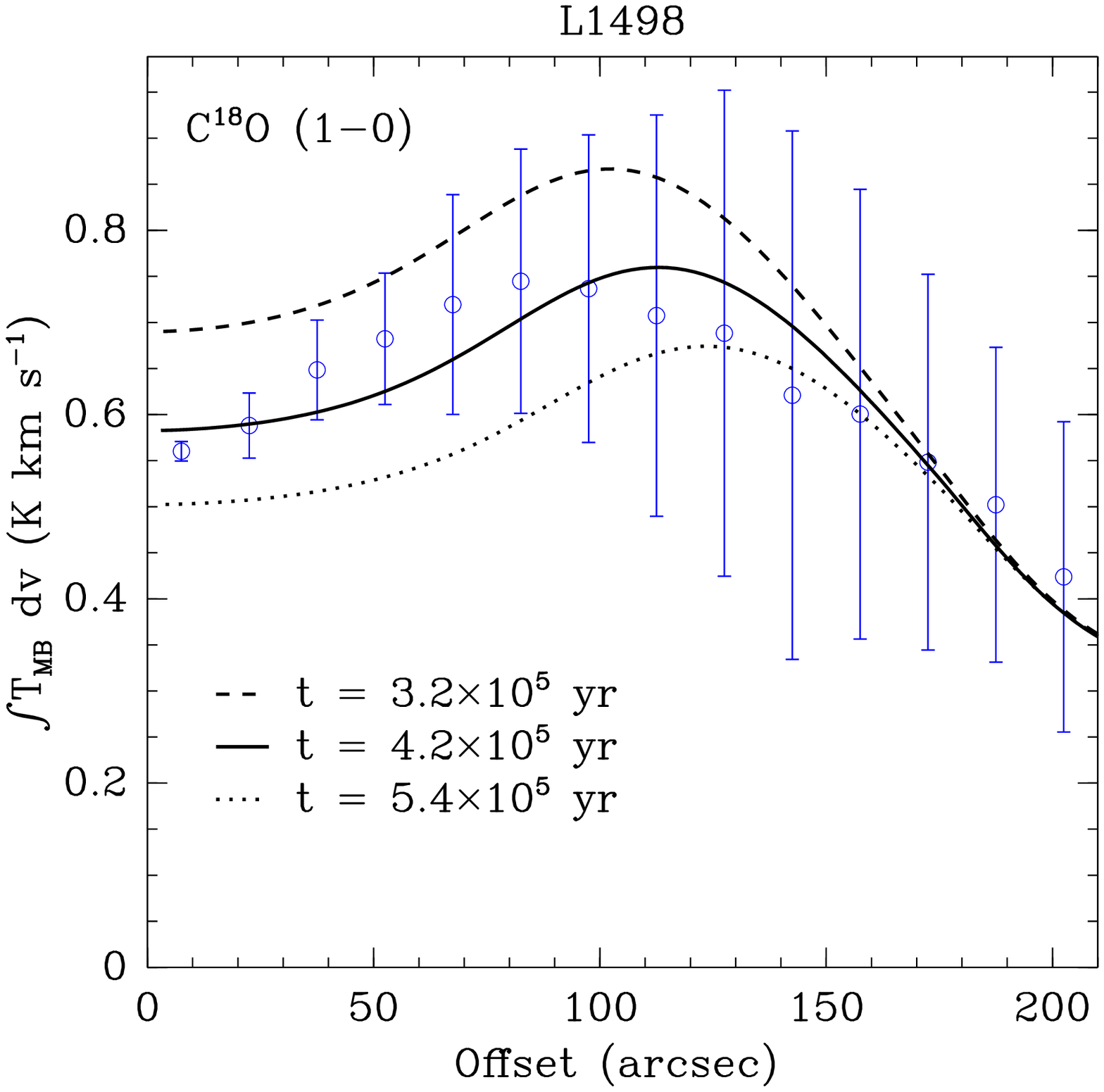} &
    \includegraphics[width=8cm]{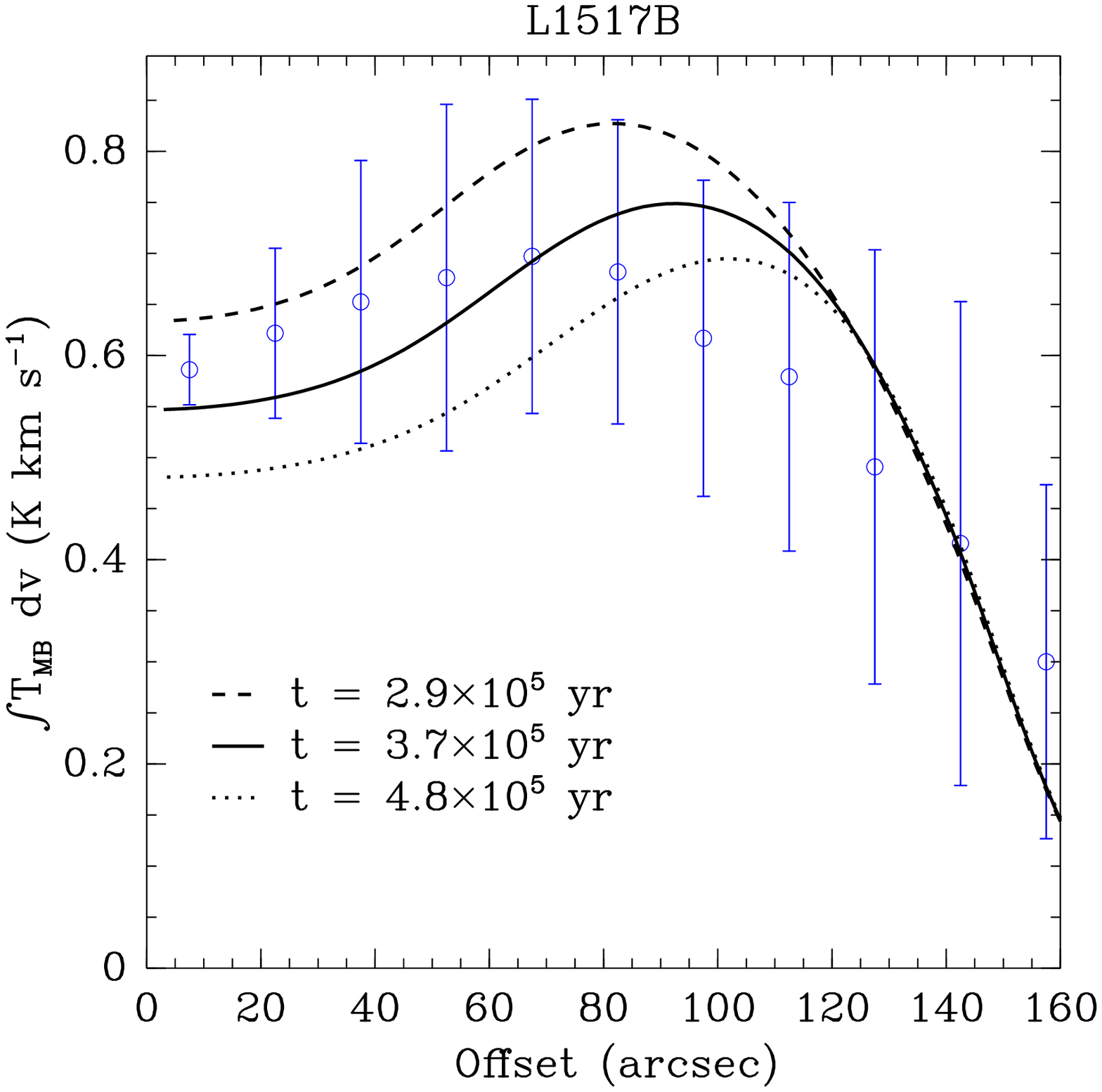} \\
  \end{tabular}
  \caption{Comparison between the observed C$^{18}$O~(1-0) \edit{integrated}
    intensity as a function of \edit{the} offset from the core center
    (blue dots with error bars) \edit{and} the predictions our
    fiducial model for different chemical ages (solid, dotted and
    dashed lines) in L1498 (left panel) and L1517B (right
    panel). Observations have been radially (or ellipsoidally)
    averaged, and the error bar\edit{s} show the 1$\sigma$ standard
    deviation in each ``bin''. }
  \label{fig:c18o-depletion}
\end{figure*}

\edit{Figure}~\ref{fig:c18o-depletion} compares the observations of
the C$^{18}$O~(1-0) line intensity with the predictions of our
model. In this figure, we plot the \edit{integrated} intensity as a
function of the offset from the dust continuum peak\footnote{The
  continuum peak offsets from the map center are (-10\arcsec ; 0) and
  (-15\arcsec ; -15\arcsec) in L1498 and L1517B, respectively.}. As
mentioned in section \ref{sec:observations}, the continuum and line
emission in L1498 is not circular, but it is elongated along the
Southeast-Northwest direction.  The C$^{18}$O~(1-0) line intensity is
therefore averaged along ellipses with an aspect ratio of 0.6, and a
position angle of -40\degr. In L1517B, the emission is almost
spherical, and the \edit{integrated} intensities are averaged
radially. In order to estimate the uncertainty on the averaged
intensity, we computed the standard deviation of the measured
\edit{values} along each ellipse (or circle). Consequently, the error
bars on the data points in Fig.~\ref{fig:c18o-depletion} reflect more
the departure from spherical (or ellipsoidal) symmetry than the
statistical noise on each data point (which is much smaller).
 
On the first panel of Fig~\ref{fig:c18o-depletion}, we see that in
L1498 the averaged C$^{18}$O~(1-0) \edit{integrated} intensity increases from
0.45~K~km~s$^{-1}$ at an offset of 200\arcsec, up to a peak value of
0.75~K~km~s$^{-1}$ at an offset of 100\arcsec. Closer to the core
center \edit{the integrated intensity} decreases to 0.55~K~km~s$^{-1}$.  This
figure also shows the predictions of our fiducial model for different
times in our chemical simulation\footnote{This time is named chemical
  age hereafter, and we discuss its the meaning in term\edit{s} of
  core physical age in \S~\ref{sec:constraints-core-age} }. This
fiducial model assumes a constant grain size of 0.1~$\mu m$, a
cosmic-ray ionization rate $\zeta = 3 \times 10^{-17}$~s$^{-1}$, and
neglects UV photodesorption from secondary photons. In addition, it
assumes \edit{that} the core is embedded in \edit{a} halo with a
density of $2 \times 10^{3}$~cm$^{-3}$, and an external radius $r_{2}
= 1.0 \times 10^{5}$~AU (0.5~pc). This halo is needed to reproduce the
extended C$^{18}$O~(1-0) emission. Although the density in the halo is
not well constrained, our modeling indicates that it \edit{cannot
  exceed} the critical density of the C$^{18}$O~(1-0) line ($2 \times
10^{\editm{3}}$~cm$^{-3}$). Models with higher densities produce too
much emission with respect to the observations at large radii ($>
150\arcsec$ in Fig.~\ref{fig:c18o-depletion}). The corresponding
visual extinction of the halo is 2.3~mag, which is consistent with the
average visual extinction observed in the region of the Taurus cloud
where L1498 is located \citep{Cambresy99}. In addition, we have
assumed that the cores are irradiated externally by a FUV field of
$1.7~G_{0}$ in \cite{Habing68} units, that is the local average
interstellar radiation field (ISRF) in the FUV. We found that the
observations are well reproduced by the fiducial model for a chemical
age of $4.2 \times 10^{5}$~yr. In particular, the model reproduces
the increase of the emission up to the observed peak at $A_v \sim 10$,
and then the decrease at higher $A_{v}$. Indeed, the observations
allow us to place a stringent constraint on the core chemical age: for
slightly smaller or greater ages ($t = 3.2 \times 10^{5}$ and $5.4
\times 10^{5}$~yr), the model produces too much or too little
emission, respectively.

The second panel of Fig~\ref{fig:c18o-depletion} presents the same
comparison for L1517B. In this source, a similar pattern as in L1498
is seen: the C$^{18}$O~(1-0) \edit{integrated} intensity increases
from 0.25~K~km~s$^{-1}$ offset of $\sim 160\arcsec$ up to a peak of
0.70~K~km~s$^{-1}$ around 70\arcsec, and then decreases down to
0.60~K~km~s$^{-1}$ at the dust continuum position. However, the
extended emission is half that of L1498. In addition, in L1517B the
C$^{18}$O~(1-0) emission peak is located closer to the dust continuum
than in L1498 (70 and 100\arcsec, respectively). Our model reproduces
the observed emission reasonably well. However, \edit{we} predict a
peak in the C$^{18}$O emission between 80 and 110\arcsec, depending on
the adopted core age, \edit{which} is somewhat larger than the
observed emission peak (70\arcsec). \edit{However, the} model allows
us to \edit{precisely} constrain the core chemical age, e.g. the
observations at offsets smaller than 100\arcsec\ are bracketed between
models with $t = 2.9 \times 10^{5}$ and $4.8 \times 10^{5}$~yr. The
model that fits the observation\edit{s} the best has $t = 3.7 \times
10^{5}$~yr. As for L1498 we have assumed that the core is embedded in
a spherical halo. The halo has the same density as in L1498 ($2 \times
10^{3}$~cm$^{-3}$) but has a smaller external radius ($r_{2} = 5.9
\times 10^{4}$~AU, 0.3~pc). This corresponds to an $A_v$ of 1.0
magnitude, again \edit{in} rough agreement with \citet{Cambresy99}
extinction maps of the region. As for L1498, we have assumed and
external ISRF of $1.7~G_{0}$.

\begin{figure*}
  \centering
  \begin{tabular}{cc}
    \includegraphics[width=8cm,angle=0]{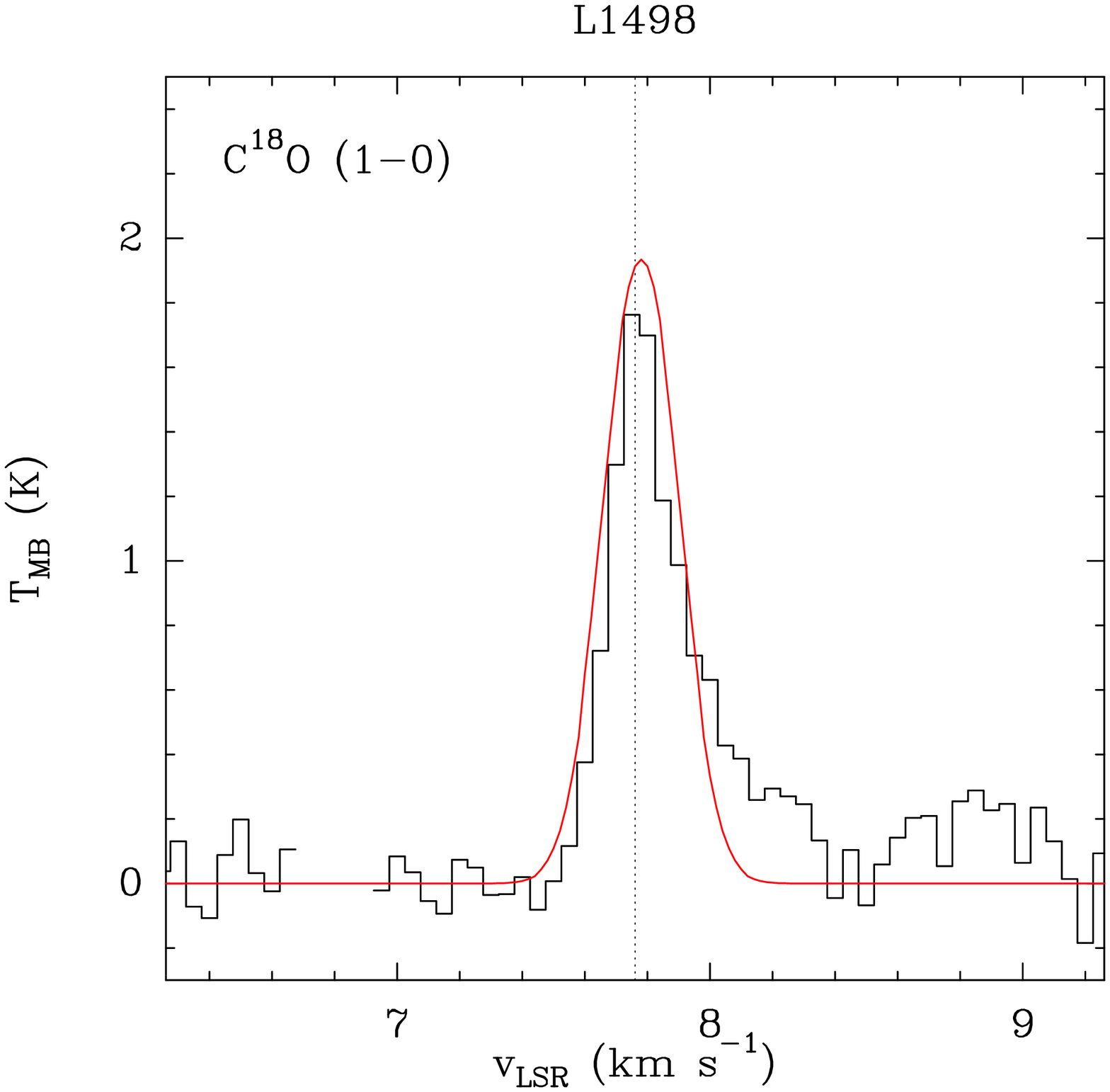} &
    \includegraphics[width=8cm,angle=0]{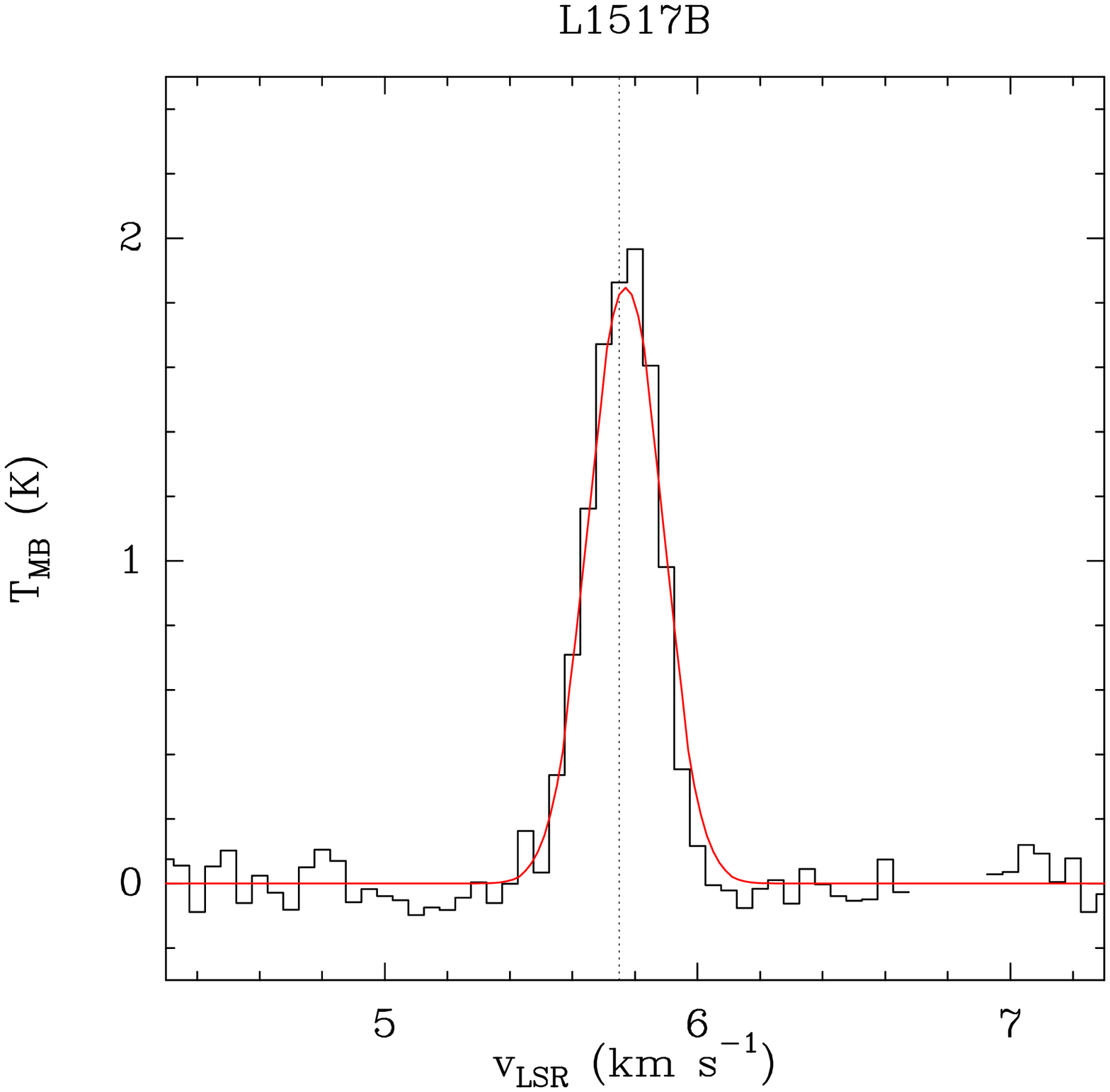} \\
  \end{tabular}
  \caption{Comparison between the observed C$^{18}$O~(1-0) line
    profile in L1498 (left panel) and L1517B (right panel) \edit{and} our
    fiducial model towards the center each source. In each panel the
    black histograms show the observation, and the red curves show
    the predicted line profile. The vertical dashed lines indicate the
    source rest velocity (7.76 and 5.75 km~S$^{-1}$ for L1498 and
    L1517B, respectively), as determined from Gaussian fits of the
    observed line profiles. Bad correlator channels around 6.8
    km~s$^{-1}$ have been masked. 
  }
  \label{fig:c18o-spec}
\end{figure*}

\edit{Figure}~\ref{fig:c18o-spec} compares the observed line profiles
with the \edit{model predictions} towards the center of L1498 and
L1517B. In L1498 the observed profile has a Gaussian shape peaked
around 7.8~km~s$^{-1}$, with a red wing extending from about 8 to
8.5~km~s$^{-1}$. However, inspection of the channel maps indicates
that this wing component has a different spatial distribution
tha\edit{n} the Gaussian component, and is probably not related to the
core \citep[see also][Appendix B]{Tafalla06}. Therefore this component
was not included in our analysis (it was also excluded from the
integrated intensities shown in Fig.~\ref{fig:c18o-depletion}). In
L1517B the observed C$^{18}$O~(1-0) line profile also has a Gaussian
shape, centered around 5.8~km~s$^{-1}$ \edit{with} no wings. The
latter is well reproduced by our model if one assumes a non-thermal
broadening of 0.23~km~s$^{-1}$ (FWHM). For the same non-thermal
broadening the line profile in L1498 is also reasonably well
reproduced, although the model somewhat overestimates the width of the
Gaussian component.

\subsection{$\mathrm{H^{13}CO^{+}}$ emission}
\label{sec:h13co+-emission}

We now attempt to model the H$^{13}$CO$^{+}$~(1-0) line observations
in the two cores. This ion is mainly formed by the reaction of
H$_{3}^{+}$ with $^{13}$CO, and is destroyed by the dissociative
recombination with free electrons. Our C$^{18}$O~(1-0) analysis places
constraints on the C$^{18}$O abundance in the core, and indirectly on
the $^{13}$CO abundance. Thus with the $^{13}$CO \edit{abundance
  structure} fixed, the H$^{13}$CO$^{+}$~(1-0) abundance depends on
both the H$_{3}^{+}$ and on the electron abundances. The H$_{3}^{+}$
abundance is set by the cosmic-ray ionization rate. The electron
abundance is set by different mechanisms, depending on the $A_{v}$. At
$A_{v} \le 5-6$, it is controlled by the photo-ionization of neutral
carbon by UV photons, while at greater extinctions, UV can not
penetrate and the gas ionization is mainly due to cosmic-rays
\cite[see ][Fig.~11]{Bergin07b}. In addition, metals with a low
ionization potential (Fe, Na, and Mg) also influence the ionization
fraction, because they recombine with electrons very slowly. \edit{The
  cosmic-ray ionization rate and the low ionization potential metal
  abundance} have similar effects on the ionization fraction, and they
are difficult to constrain independently from our
H$^{13}$CO$^{+}$~(1-0) observations alone. In B68 we found a metal
abundance of $3 \times 10^{-9}$, but we suggested that the
observations were also consistent with a full depletion of metals
\citep{Maret07a}. Here we adopt the same metal abundance as in B68,
and we adjust the cosmic-ray ionization rate in order to reproduce the
observations.

\subsubsection{Effect of the cosmic-ray ionization rate}

\begin{figure*}
  \centering
  \begin{tabular}{cc}
    \includegraphics[width=8cm]{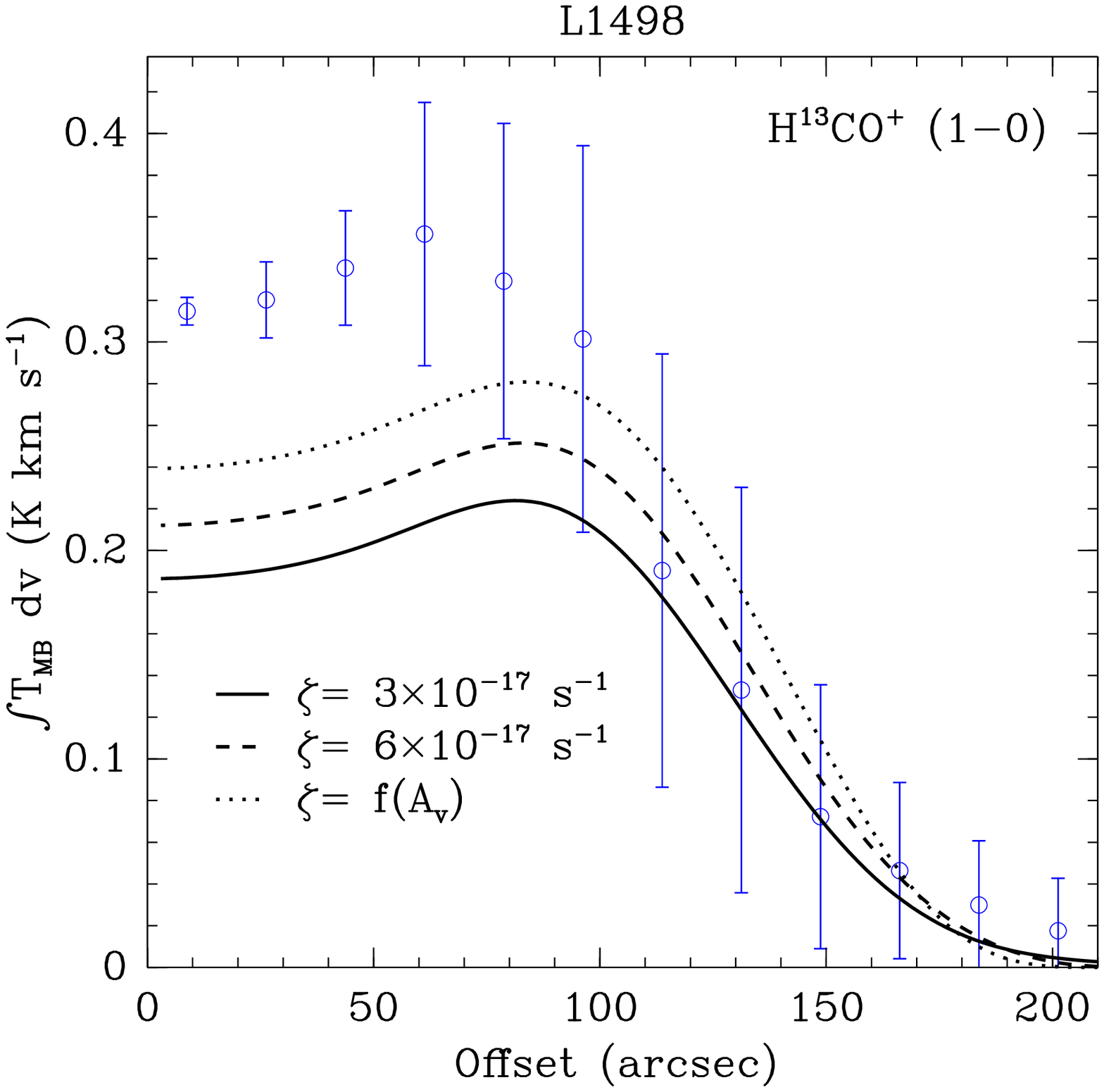} &
    \includegraphics[width=8cm]{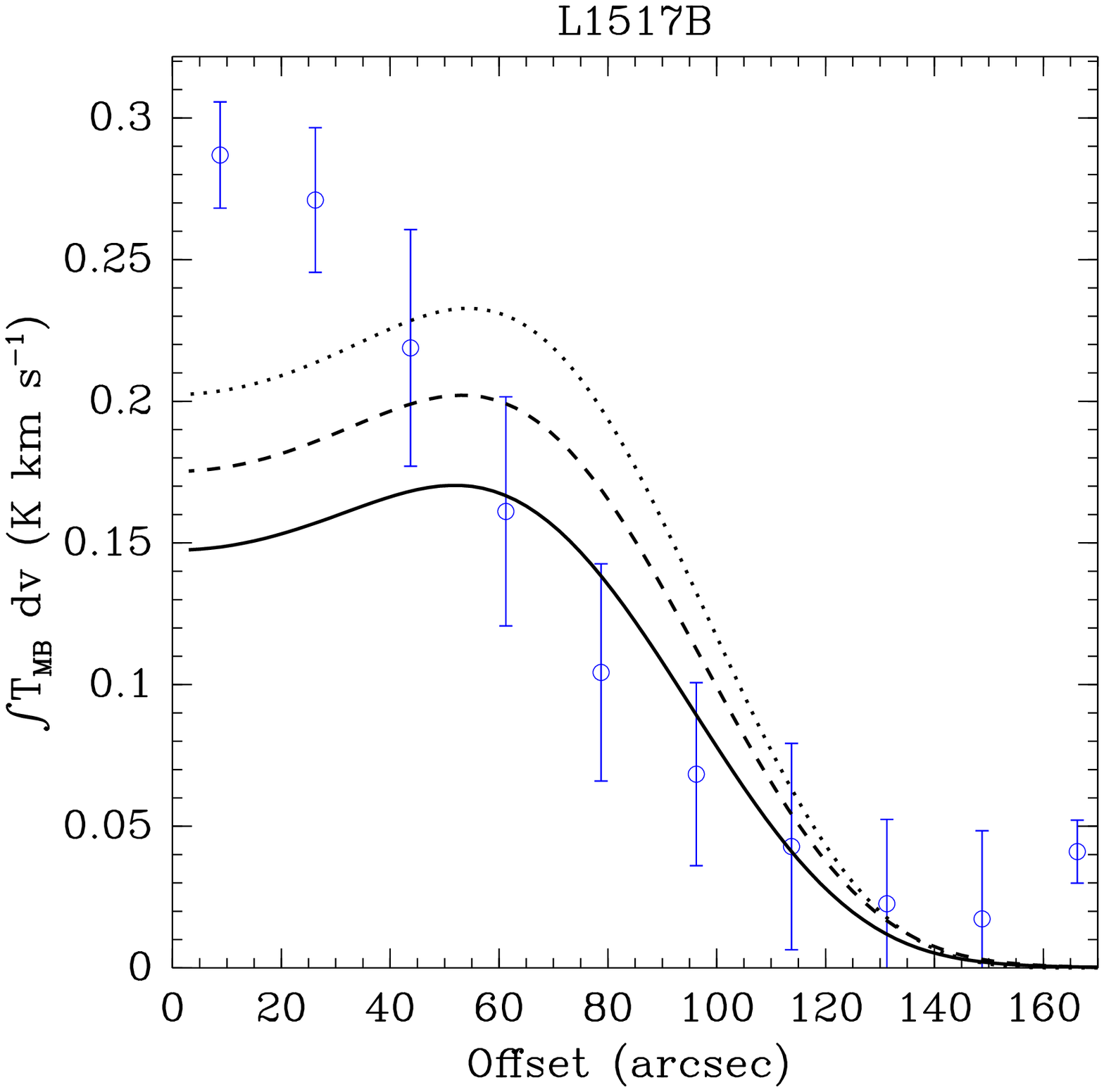} \\
  \end{tabular}
  \caption{Comparison between the observed H$^{13}$CO$^{+}$~(1-0)
    \edit{integrated} intensity as a function of \edit{the} offset
    from the core center with the predictions of our model for
    different cosmic-ray ionization rates (solid, dotted and dashed
    lines) in L1498 (left panel) and L1517B (right panel). The solid
    curve\edit{s} correspond to the fiducial model\edit{s} ($\zeta = 3
    \times 10^{-17}$~s$^{-1}$). The dashed curves (labeled $\zeta =
    f(A_v)$) correspond to model predictions for a variable cosmic
    ionization rate as a function of $A_v$ (see text).}
  \label{fig:h13cop-cosmic-ray}
\end{figure*}

\edit{Figure}~\ref{fig:h13cop-cosmic-ray} compares the observed
H$^{13}$CO$^{+}$~(1-0) \edit{integrated} emission in L1498 and L1517B
with the predictions of our model for different cosmic-ray ionization
rates $\zeta$. Three different values of $\zeta$ are considered: $3
\times 10^{-17}$~s$^{-1}$, $6 \times 10^{-17}$~s$^{-1}$, and a
variable cosmic-ionization rate as a function of the $A_{v}$. For the
latter, we use the model of \citet{Padovani09}, who studied the
propagation of cosmic-ray protons and electrons in a molecular cloud,
for various incident electron and proton energy spectra. Here we use
their results assuming the proton energy spectrum from
\cite{Moskalenko02}, and an electron energy spectrum from
\citet[][model C]{Strong00}. This model, labeled ``M02+C00'' \edit{by}
\citeauthor{Padovani09}, was found to provide a good agreement with
the observed values of $\zeta$ for $N(\mathrm{H_{2}})$ between
10$^{20}$ and 10$^{21}$~cm$^{-2}$ (see their Fig.~15).

In L1498, all three models are in good agreement with the
H$^{13}$CO$^{+}$~(1-0) line observations for offsets greater than
100\arcsec. However, all models fail to reproduce the emission towards
the core center. The model with $\zeta = 6 \times 10^{-17}$~s$^{-1}$
and the model with a\edit{n} $A_{v}$ dependent cosmic-ray ionization
rate give a slightly better \edit{observational} agreement towards the
core center, but still significantly underproduce the emission. In
addition, these two models where found to underproduce the extended
C$^{18}$O~(1-0) emission (which is \edit{matched at} $\zeta = 3 \times
10^{-17}$~s$^{-1}$, see Fig.~\ref{fig:c18o-depletion}). This is
because CO and its isotopologues are efficiently destroyed by the
reaction with He$^{+}$ in the halo, from where the extended
C$^{18}$O~(1-0) emission arises.  He$^{+}$ itself is produced by
cosmic-ray ionization, so it\edit{s} abundance increases as the
cosmic-ionization rate increases.

In L1517B, only the model with $\zeta = 3 \times 10^{-17}$~s$^{-1}$
match\edit{es} the \edit{H$^{13}$CO$^{+}$~(1-0)} observations for
offsets greater than 50\arcsec; other models predict a more extended
emission than what is observed. Like in L1498, all models somewhat
underproduce the emission at the core center. In fact, our models
predicts a decrease in the H$^{13}$CO$^{+}$~(1-0) emission towards the
center of the core that is not observed: the H$^{13}$CO$^{+}$~(1-0)
emission in this source peaks at the center of the core (see
Fig~\ref{fig:l1517b-map}). In the following, we adopt a $\zeta$ of $3
\times 10^{-17}$~s$^{-1}$, which provides the best-fit to the
C$^{18}$O~(1-0) and H$^{13}$CO$^{+}$~(1-0) emission in both cores, and
we investigate different mechanisms to explain the larger
H$^{13}$CO$^{+}$~(1-0) intensities observed towards the center of the
two cores. This value of $\zeta$ is in good agreement with that
measured in other cores
\citep[$10^{-18}-10^{-16}$~s$^{-1}$]{Caselli98} including B68
\citep[$(1-6) \times 10^{-17}$~s$^{-1}$; ][]{Maret07b}.

\subsubsection{Effect of cosmic-ray induced photodesorption}

Although our C$^{18}$O~(1-0) \edit{data} places constraints on the CO
abundance in the core, this line has a relatively low critical density
($2 \times 10^{\editm{3}}$~cm$^{-3}$ at 10~K) and is therefore
probing mostly the outer \edit{layers}. On the other hand,
H$^{13}$CO$^{+}$~(1-0) has a critical density which is about \edit{two
  orders} of magnitude higher ($1.5 \times 10^{5}$~cm$^{-3}$) and
\edit{likely traces dense material deeper in the cores} (the critical
density of this line is comparable to the central densities of the two
cores, see Table~\ref{tab:density-profiles}). Therefore, the excess of
H$^{13}$CO$^{+}$~(1-0) emission observed in the two cores with respect
to our model predictions hints that our model underestimate\edit{s}
the abundance of CO and its isotopologues in the inner parts of the
core, and this underestimate propagates into a lower-than-observed
H$^{13}$CO$^{+}$~(1-0) intensity. In these regions, CO is removed from
the gas phase by the freeze-out on the dust grains. Thermal
evaporation being inefficient a 10~K, desorption mostly occurs through
cosmic-rays.

\begin{figure*}
  \centering
  \begin{tabular}{cc}
    \includegraphics[width=8cm]{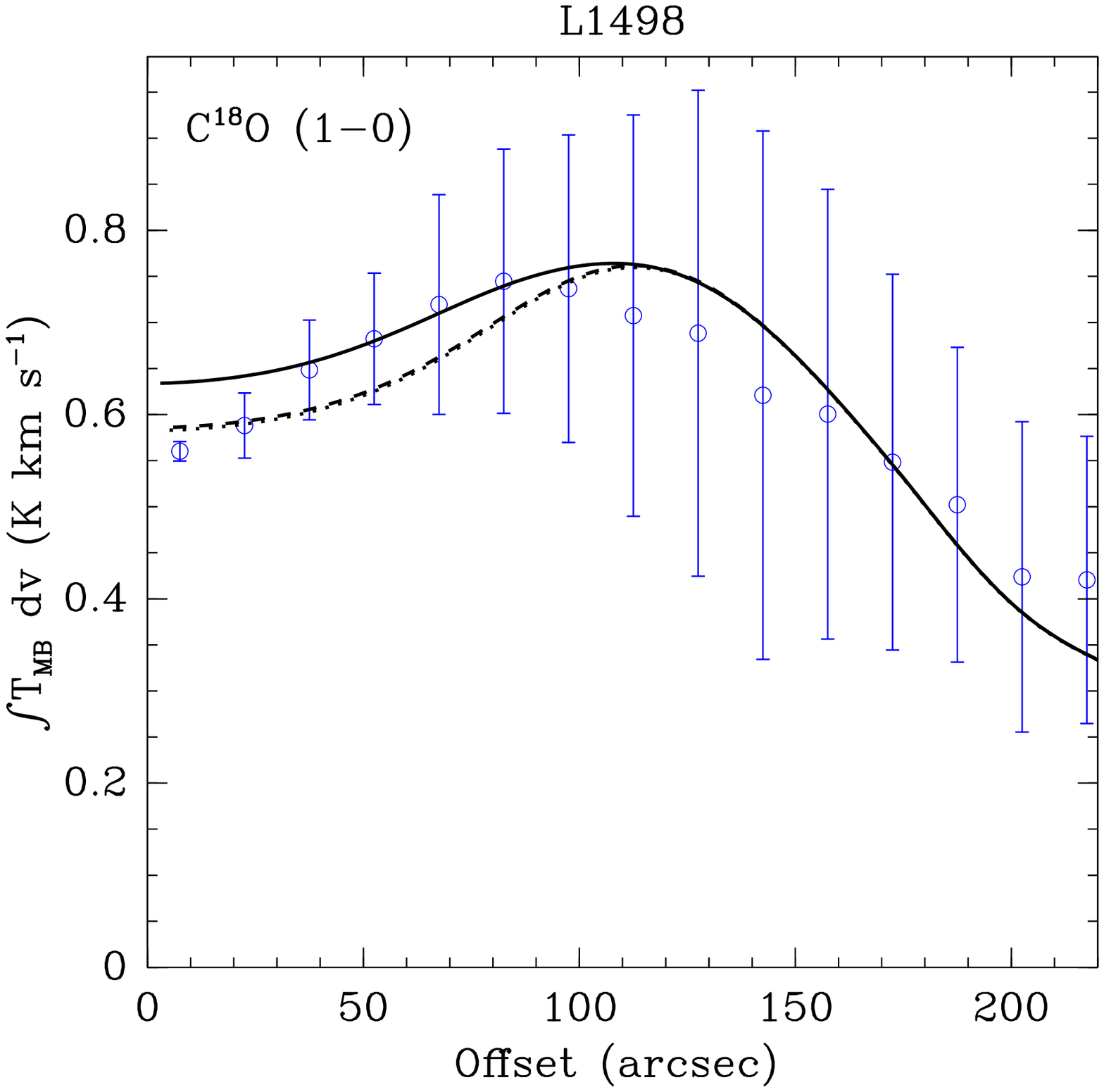} &
    \includegraphics[width=8cm]{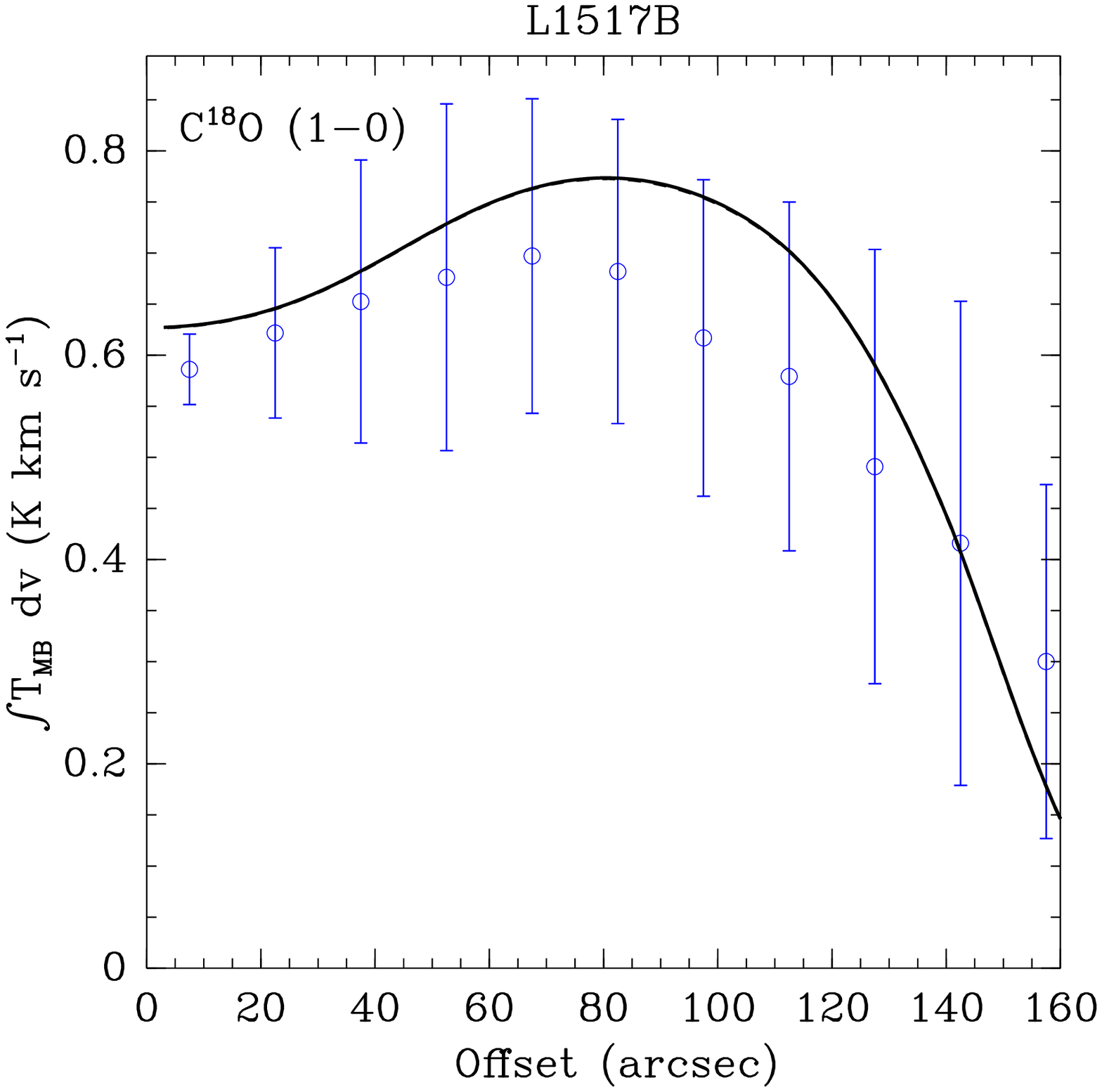} \\
    \includegraphics[width=8cm]{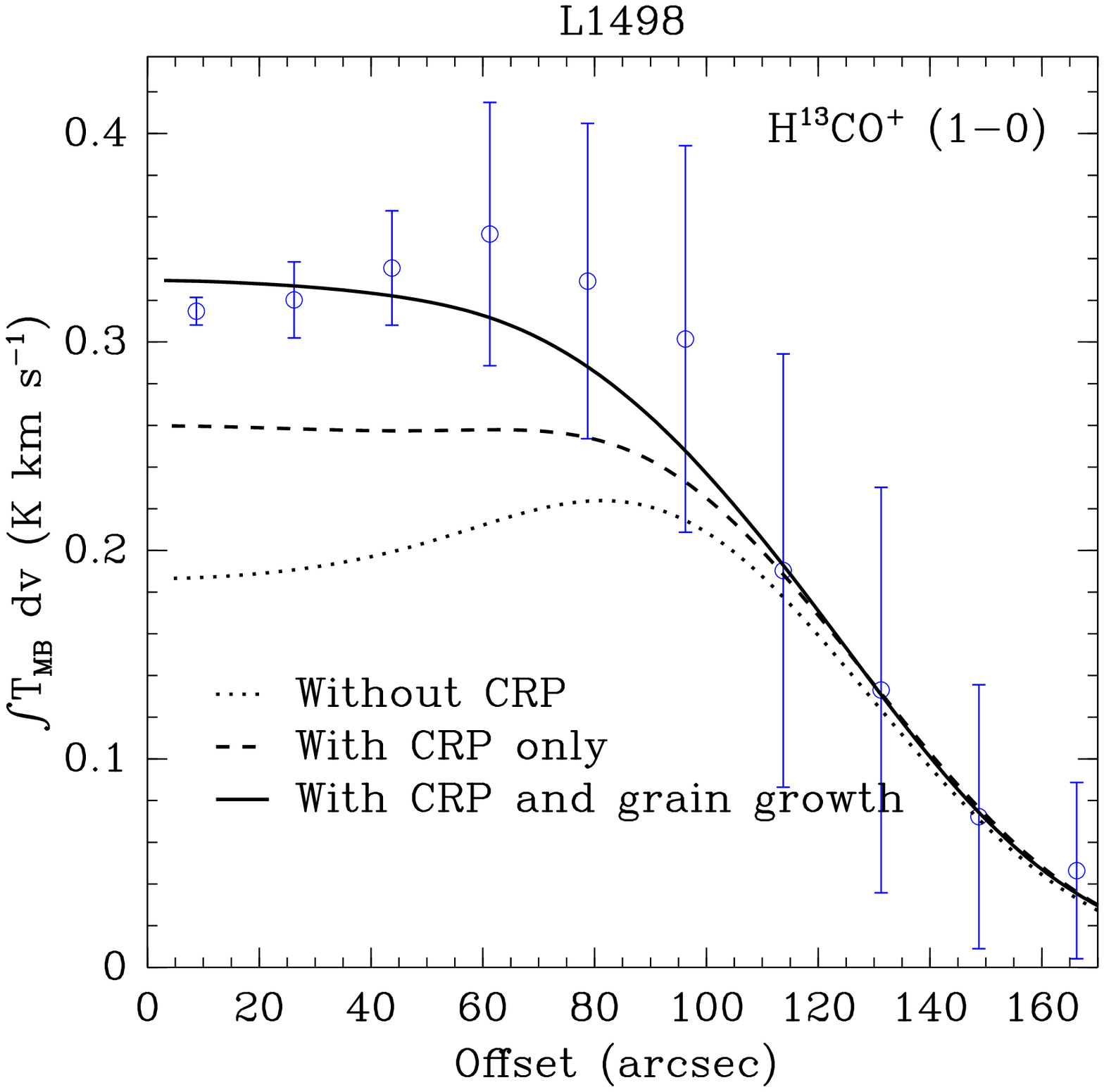} &
    \includegraphics[width=8cm]{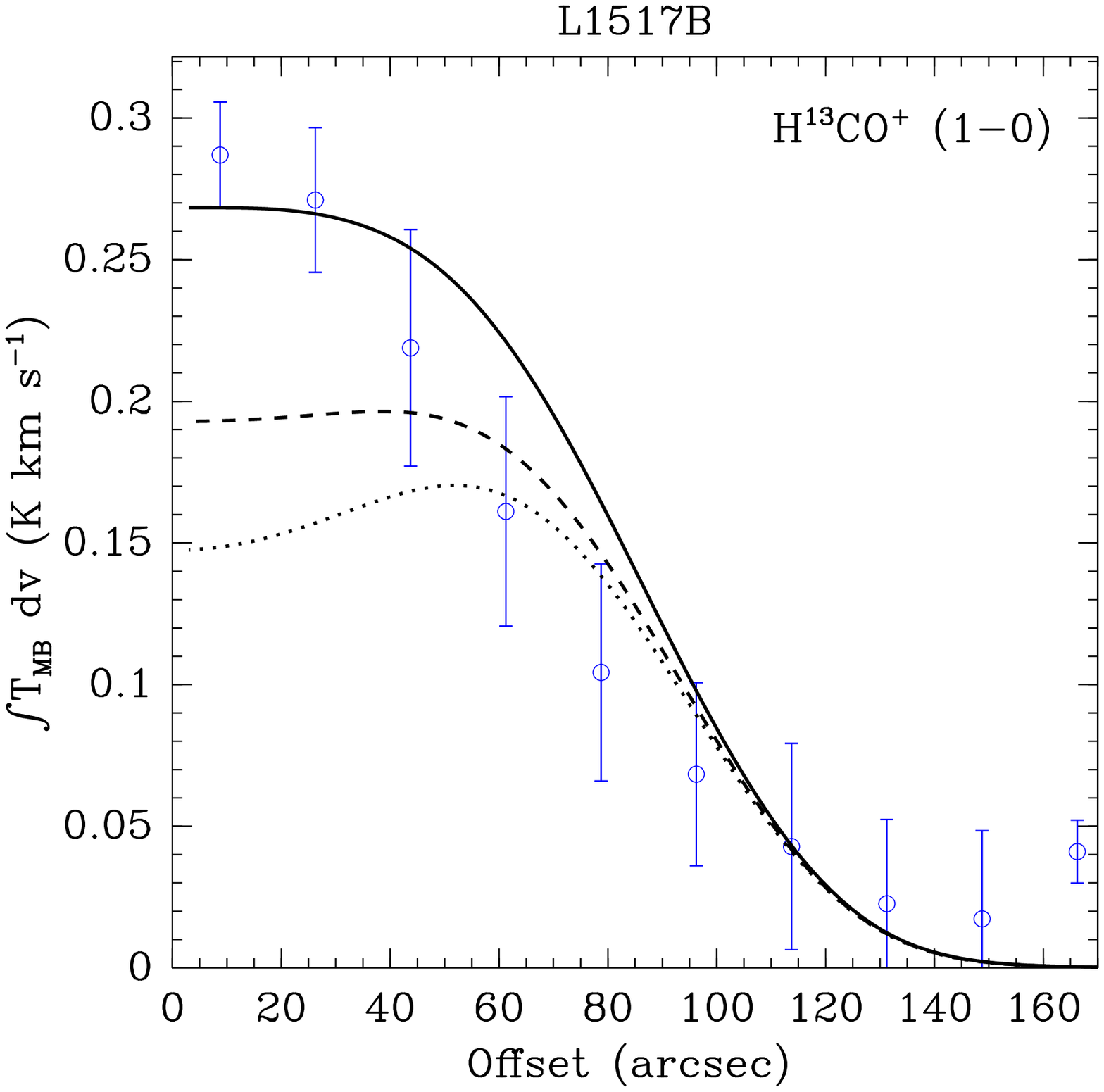} \\
  \end{tabular}
  \caption{Comparison between the observed C$^{18}$O~(1-0) (top
    panels) and H$^{13}$CO$^{+}$~(1-0) (bottom panels) \edit{integrated}
    intensities as a function of \edit{the} offset from core center
    \edit{and} the predictions of our models including cosmic-ray
    photodesorption (CRP) and grain-growth. In each panel the dotted
    curve corresponds to the fiducial model (no CRP and \edit{a}
    constant grain size), the dashed curve corresponds to the model
    with CRP only, and the solid curve correspond\edit{s} to the best-fit
    model (with both CRP and grain-growth). \label{fig:h13cop-crp-gg}}
\end{figure*}

Our fiducial models only include direct cosmic-ray desorption,
i.e. desorption caused by impact of the cosmic-ray particle on
grains. However, cosmic-ray desorption can also occurs in indirect
fashion: the ionization of H$_{2}$ followed by electronic
recombination, producing a UV photon. This cosmic-ray induced UV
radiation field can in turn photodesorb ices \citep[see
e.g.][]{Caselli12}. Following \citep{Oberg09b}, we have included this
effect in our model by adding a cosmic-ray induced UV field of
10$^{-4}~G_{0}$ \citep{Shen04} to that produced by the
ISRF. \edit{Figure}~\ref{fig:h13cop-crp-gg} compares the models
predictions with and without cosmic-ray photodesorption to both the
C$^{18}$O~(1-0) and H$^{13}$CO$^{+}$~(1-0) integrated
intensities. Models with cosmic-ray photodesorption produce more
H$^{13}$CO$^{+}$~(1-0) towards the core center and are in better
agreement with the observations. However, they still do not produce
enough emission with respect to the observations, suggesting that
another process is at play. In the following we explore some solutions
to this discrepancy.

\subsubsection{Effect of grain growth}

Another possible explanation for the excess of H$^{13}$CO$^{+}$~(1-0)
emission in both cores is that the grain size varies as a function of
the radius. If grain\edit{s} have coagulated at the center of the
cores, the average grain cross-section decreases, and the depletion
may be less efficient. For our fiducial models we have assumed that
grain\edit{s} are spherical and have a constant 0.1$~\mu m$ size
throughout the cores. However, both L1498 and L1517B appear in
emission at 3.6$~\mu m$ in Spitzer maps, a phenomenon know as
core-shine \citep{Pagani10}. This indicates that grains have grown
from their average 0.1$~\mu m$ size to a bigger size, up to about a
$1~\mu m$. Here we investigate the effect of grain growth on the
predicted H$^{13}$CO$^{+}$~(1-0) emission.

\citet{Steinacker10} modeled the mid-infrared emission from
\object{L183} observed with Spitzer, and found a good agreement
between the model and the observations for a parametric grain size of
the form:

\begin{eqnarray}
    a(r) & = & a_{0} \left( \frac{n(r)}{n_{0}} \right)^\alpha \quad n(r) > n_{0}\\
    & = & a_{0} \qquad \qquad \; n(r) \le n_{0}
\end{eqnarray}

Following this approach we \edit{compute} several models with a
parametric grain size. To enforce mass conservation, the grain
abundance \edit{is} recomputed at each radius, assuming the same grain
composition (i.e. olivine) and gas-to-dust mass ratio (100) \edit{as
  in} our fiducial model\footnote{\edit{In practice, this assumes that
    grains have coagulated and that the aggregates have compacted into
    a spherical shape, resulting in a lower cross-section per H
    nucleus. However, grain aggregates may have a ``fluffy'' shape, and
    their cross-section per H nucleus may be larger than what is
    assumed here.}}. The parameters for grain size \edit{are then}
adjusted until a good match with the observations was
found. \edit{Figure}~\ref{fig:h13cop-crp-gg} shows the best fit model
with both cosmic-ray photodesorption and grain growth included, for
the parameters given in Table~\ref{tab:grain-size-params}. This model
provides a good fit to the H$^{13}$CO$^{+}$~(1-0) emission, especially
at low radii. Thus, inclusion of both cosmic-ray photodesorption and
grain growth in the model appears to be \edit{the} only way to
increase the CO abundance at the core center. We note, however, that
the grain size parameters are not unique: comparable fits \edit{are}
found for larger values of $n_{0}$ and $\alpha$ (i.e. smaller
coagulation regions with steeper size decrease with the density). On
the same \edit{F}igure we show the model predictions for the
C$^{18}$O~(1-0) emission. The grain growth is found to have little
effect on this line, because it originates from regions with density
below $n_{0}$.

\begin{table}
  \caption{Grain size parameters}
  \label{tab:grain-size-params}
  \centering
  \begin{tabular}{lll}
    \hline
    \hline
    & L1498 & L1517B\\
    \hline
    $n_{0}$ (cm$^{-3}$) & $\editm{1.75 \times 10^{4}}$ & $\editm{1.25 \times 10^{4}}$\\
    $a_{0}$ ($\mu$m) & 0.1 & 0.1 \\
    $\alpha$ & 0.25 & 0.25\\
    \hline
  \end{tabular}
\end{table}

One possible inconsistency of our approach is that the density profiles
in the two cores have been derived assuming a typical dust opacity at
1.2~mm \citep[0.5~cm$^{2}$~g$_\mathrm{dust}^{-1}$;][]{Tafalla02}. In
principle, grain growth could affect this opacity, and in turn the
density we derive for the two cores. \citet{Ormel11} computed the
effects of grain coagulation on the dust opacity, and found that for
$t \le 3 \times 10^{6}$~years, the change in the opacity at 850~$\mu$m
are rather small (a factor 2 at most depending on the grain
composition, see their Fig.~6). We conclude that grain growth should
not affect our density determinations, and that our approach is
consistent.

\begin{figure*}
  \centering
  \begin{tabular}{cc}
    \includegraphics[width=8cm,angle=0]{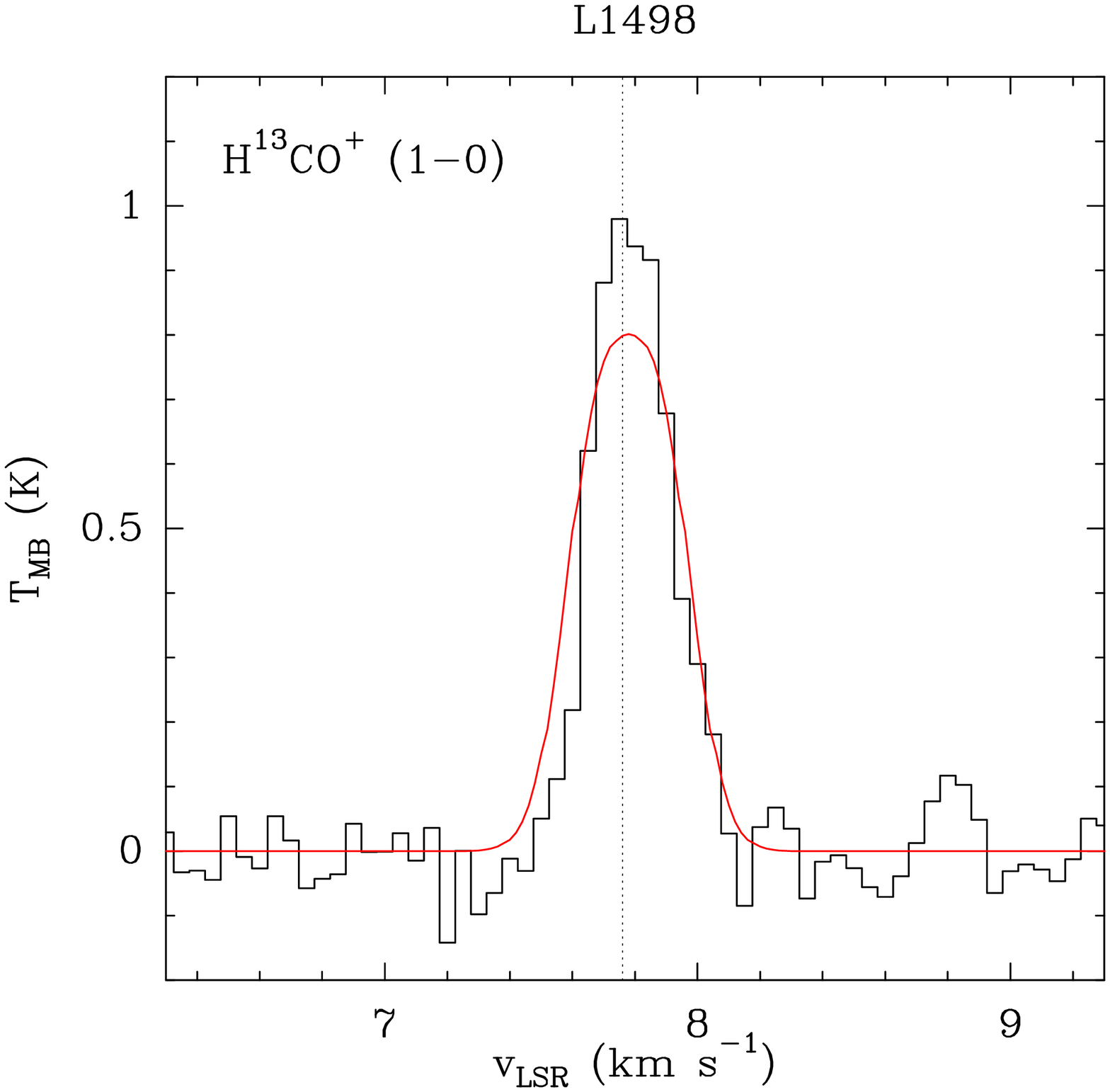} &
    \includegraphics[width=8cm,angle=0]{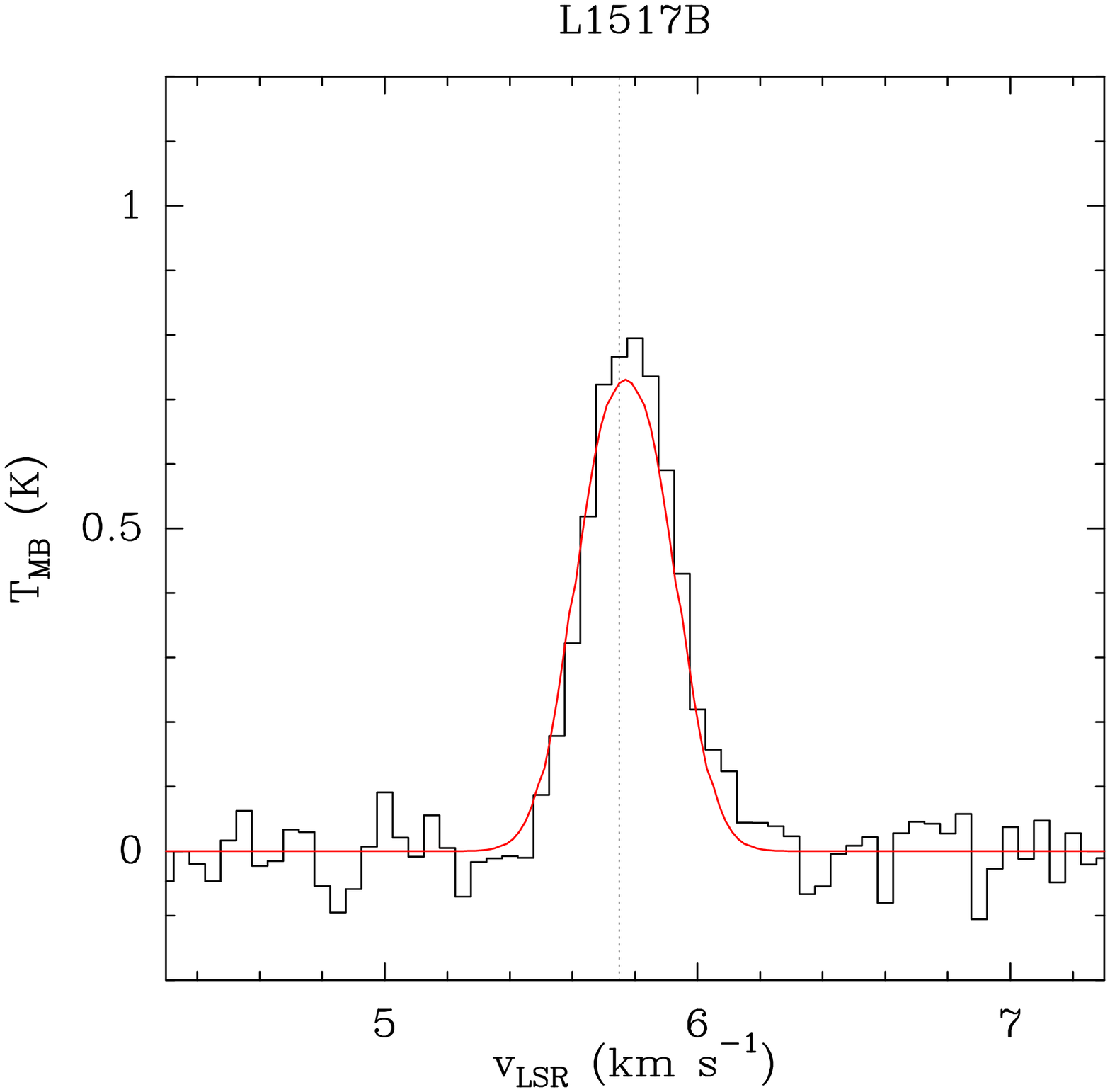} \\
  \end{tabular}
  \caption{Comparison between the observed H$^{13}$CO$^{+}$~(1-0) line
    profiles in L1498 (left panel) and L1517B (right panel) \edit{and} our
    best-fit models towards the center each source. 
    \label{fig:h13cop-spec}}
\end{figure*}

Finally, Fig.~\ref{fig:h13cop-spec} compares the observed profiles
towards the core center with the model predictions for our best fit
model (i.e. models including both cosmic-ray desorption and grain
growth). In both cores, the line profiles are close to Gaussians, and
are well reproduced by the model, assuming the same turbulent
broadening as for the C$^{18}$O~(1-0) modeling.

\section{Discussion}
\label{sec:discussion}

\subsection{Constraints on the core age}
\label{sec:constraints-core-age}

Our model and observations of the C$^{18}$O emission allow us to place
stringent constraints on the chemical age of the two cores. It is
interesting to note that this age is comparable for the two cores:
$(3-5) \times 10^{5}$~yr. For B68, \citet{Bergin06b} obtained a
somewhat younger age: $10^{5}$~yr. Here we discuss the meaning of
these chemical ages.

As discussed by \cite{Bergin06b}, the age derived from the amount of
CO depletion \edit{is} likely \edit{a} lower limit on the age of the
core, and this for two reasons. First, our model is ``pseudo-time
dependent'', in the sense that it does not includes the physical
evolution of the cloud and the core. In our \edit{modeling} we have
assumed that the density and temperature profile of the core do not
evolve with time; in reality, as the cloud condenses and fragments to
form clumps and cores, the gas density should increase by several
orders of magnitudes, while the temperature should drop as a result of
reduced UV penetration in the cloud. Therefore the cores density must
have been lower in the past, while their temperature must have been
slightly higher. Because the depletion time scale is proportional to
the gas density, and the thermal desorption time scale depends
exponentially \edit{on} the dust temperature, the chemical age derived
from CO depletion is a lower limit to the core ages.

Second, our chemical simulation starts with an initial composition
which is typical of dense molecular gas. We have assumed that H$_{2}$,
CO, H$_{2}$O (in form of ices) and a small fraction of N$_{2}$ are
pre-existing \edit{in the cloud}, with other elements assumed to be in
atomic form. Detailed photo-dissociation models show that for an H
density of $2 \times 10^{4}$~cm$^{-3}$, the transition between C$^{+}$
and CO occurs at $A_{v} \sim 0.2$~mag. In addition,
\cite{Hollenbach09} models for the formation of H$_{2}$O shows that
for $A_{v}$ greater than 4~mag, water ices are the dominant oxygen
carrier. Note that \cite{Hollenbach09} models assumes a FUV field of
100~$G_{0}$; for a FUV field of 1~$G_{0}$, the transition would occur
at a lower $A_{v}$. Thus the initial conditions adopted here are
representative of a cloud with \edit{an} extinction of 2-4~mag. In
reality the core would form from diffuse atomic gas, and the time
needed to form CO and H$_{2}$O is not included in our estimate, and
the chemical time is \edit{again} a lower limit to the core physical
ages.

The lower limits we obtain on the core ages may be compared to the
lifetime obtained from large scale surveys. \citet{Enoch08} carried
out a complete census if the prestellar cores in Perseus, Serpens and
Ophiuchus clouds (but not Taurus-Auriga) from millimeter continuum
observations, and derived an average lifetime of $4.3 \times
10^{5}$~yr, once the core density exceed their detection threshold of
$2 \times 10^{4}$~cm$^{-3}$. From optical observations, \cite{Lee99}
obtained a slightly larger lifetime $0.3-1.6 \times 10^{6}$~yr for
more diffuse cores (including Taurus-Auriga cores). These numbers
refer to the average lifetime of the cores once they reach the
detection threshold; they can therefore be considered as an upper
limit on the age of both L1498 and L1517B. Combining these upper
limits with the lower limits we obtain form the chemical ages, we
found that the two cores ages are comprised between 0.3 and
1.6~Myr. In addition, we note that the lower limits in the age we
obtain here are comparable to shortest lifetime obtained by
\cite{Lee99} ($0.3 \times 10^{6}$~yr); it is therefore possible that
two cores are at a relatively evolved stage, and may quickly evolve to
form protostars. It is interesting to note that, using an independant
technique, \cite{Pagani13} obtained an age of $<$0.7 My for the L183
core, based on its observed N$_{2}$D$^{+}$/N$_{2}$H$^{+}$ ratio; this
value is consistent with the chemical ages derived in this study.

\subsection{Core chemical structure}
\label{sec:cores-chem-struct}

\begin{figure*}
  \centering
  \begin{tabular}{cc}
    \includegraphics[width=8cm]{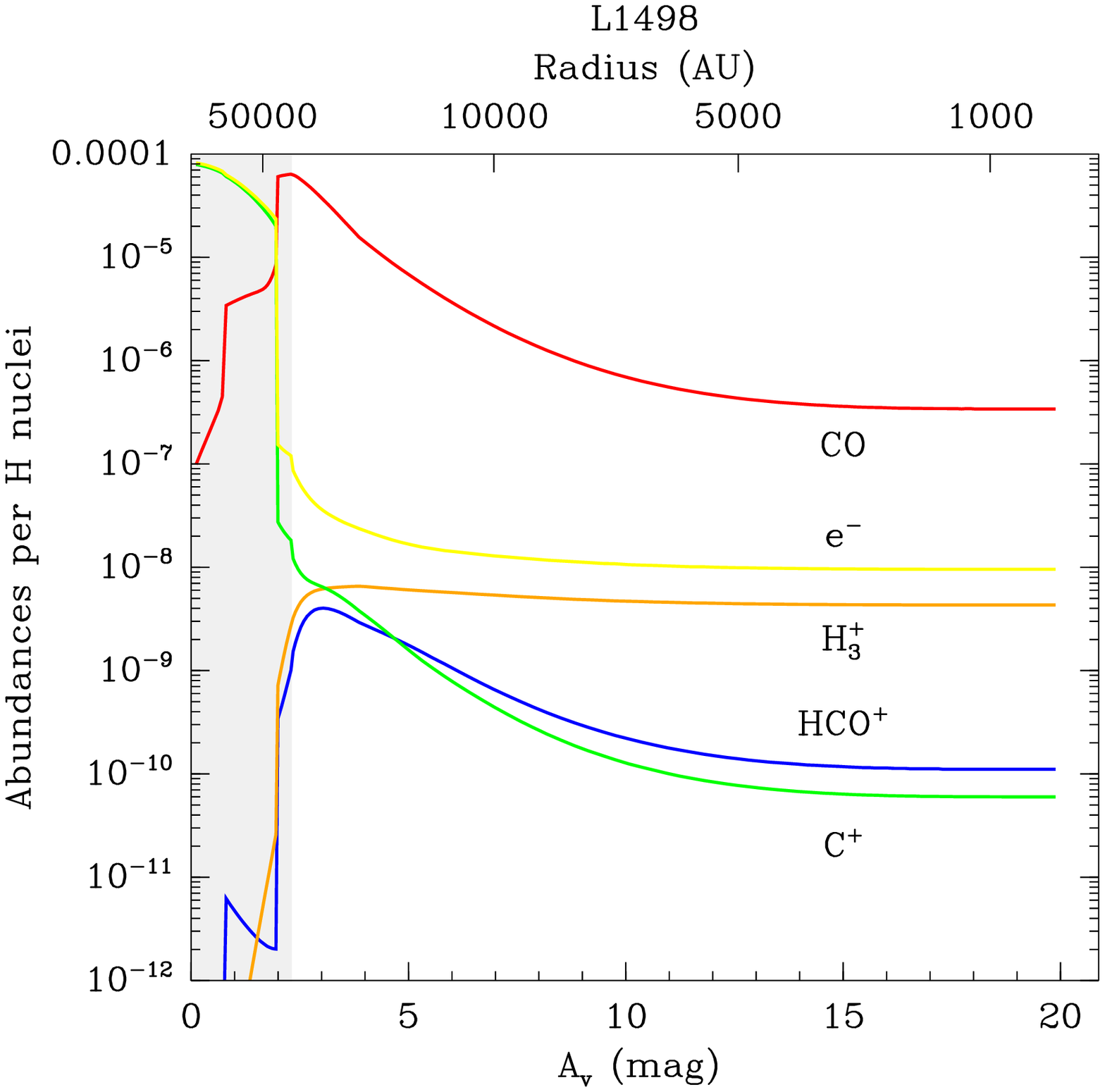} &
    \includegraphics[width=8cm]{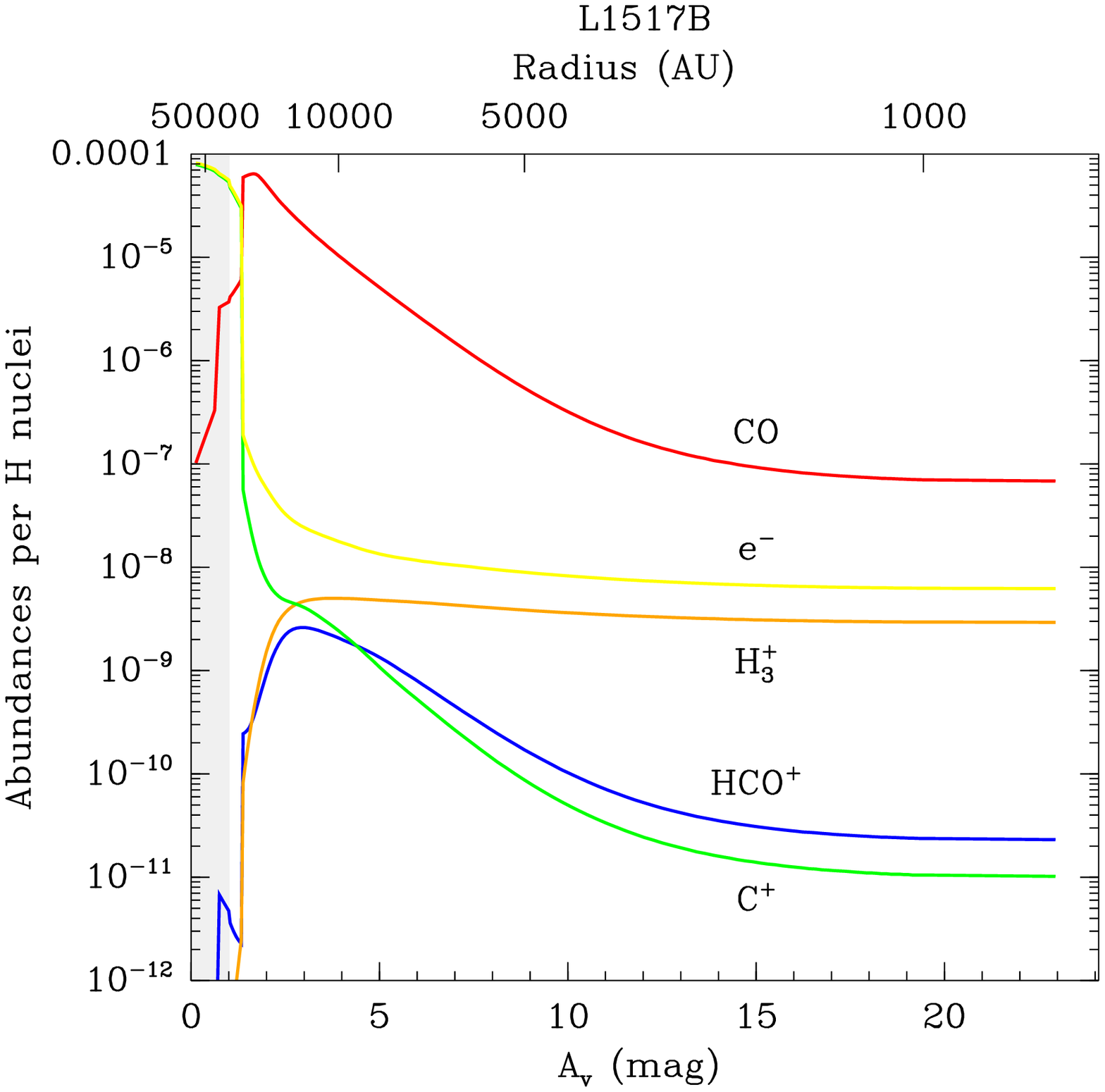} \\
  \end{tabular}
  \caption{Predicted abundances for several species as a function of
    the visual extinction and core radius in L1498 (left panel) and
    L1517B (right panel). The shaded area indicates the halo around
    each core. }
  \label{fig:abundances}
\end{figure*}

\edit{Figure} \ref{fig:abundances} shows the predicted abundances for
several species, as a function of the radius, for our best-fit
models. From the chemical point of view, the cores can be divided in
three distinct zones. At $ 0 \le A_{v} \le 1-2$, the chemistry is
dominated by photo-processes: CO is photo-dissociated, and $C^{+}$ is
the main carbon reservoir. Deeper in the cores, at $1-2 \le A_{v} \le
5$, the external UV radiation field is attenuated, and
photo-dissociation is less important. On the other hand, the density
at $A_{v} = 3$ is only $2-4 \times 10^{4}$~cm$^{-3}$, and the CO
depletion timescale \citep[see e.g.][]{Bergin95} is $0.5-1 \times
10^{5}$~yr i.e. comparable to the derived core chemical ages. As a
consequence, the CO depletion in this region is moderate, and the CO
abundance peaks at $A_{v} = 1-2$. HCO$^{+}$ -- whose parent molecule
is CO, also peaks at the same $A_{v}$. Finally, at $A_{v} > 5$, the
chemistry is dominated by the depletion heavy \edit{species}. Our
models predict a CO abundance at the core centers of $0.7-3 \times
10^{-7}$ with respect to H nuclei; this is between 2 and 3 orders of
magnitude lower than the peak CO abundance. HCO$^{+}$ is also affected
by depletion, but its abundance does not decrease as steeply as
CO. This is because HCO$^{+}$ is mainly destroyed by electronic
recombination, and the electron abundance decreases sharply towards
the core center, as discussed below. The fact HCO$^{+}$ is less
affected by depletion than CO is one of the reason\edit{s} why
H$^{13}$CO$^{+}$~(1-0) is probes deeper in the core than
C$^{18}$O~(1-0); the second reason is that the lines have different
critical densities, as already discussed.

In L1498, the ionization fraction -- which is equal to the electron
abundance because the abundance of other negative charge carriers,
i.e. anions and negatively charged grains, is negligible -- ranges
from $8 \times 10^{-5}$ in the core halo down to $1 \times 10^{-8}$ at
the core center. In L1517B, the ionization fraction ranges from $8
\times 10^{-5}$ at the surface down to $7 \times 10^{-9}$ at the
center. In the core halos (i.e. at $A_{v} < 1-2$), \edit{the}
ionization is mainly due to the photo-ionization of atomic carbon by
UV photons. Deeper in the cores, the ionization is mainly caused by
cosmic-rays.  The ionization measured at the center B68 compares well
with that of L1498 and L1517B \citep[$5 \times
10^{-9}$][]{Maret07a}. Our models show that at $A_{v} > 1-2$, the main
positive charge carriers are the H$_{3}^{+}$ and N$_{2}$H$^{+}$
ions. In both cores, H$_{3}^{+}$ is the most abundant ion and accounts
for about 50-60\% of the total charge. N$_{2}$H$^{+}$ is less abundant
and accounts for about 20-30\% of the charge \edit{(see
  Appendix~\ref{sec:model-pred-nitr} for a discussion of our model
  predictions for nitrogen bearing species)}. Other important charge
carriers are He$^{+}$ and H$_{3}$O$^{+}$, which both contribute to
about $\sim$5\% of the total charge. In the present study, we have not
considered the deuterated isotopologues of H$_{3}^{+}$
(i.e. H$_{2}$D$^{+}$, D$_{2}$H$^{+}$ and D$_{3}^{+}$). In the dense
part of cores, where CO is depleted, the abundance of these
isotopologues are expected to be enhanced with respect of H$_{3}^{+}$.

\subsection{The influence of grain growth on the chemistry}

\begin{figure}
  \centering \includegraphics[width=8cm]{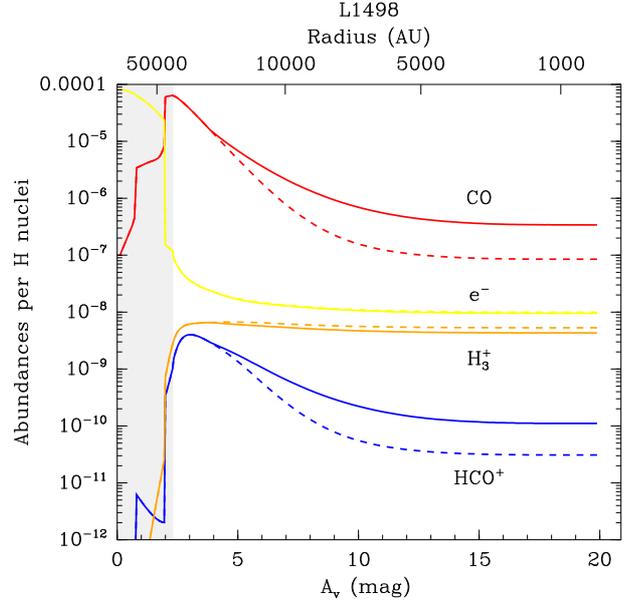}
  \caption{Comparison between the predicted abundances of several
    species in L1498 with (solid lines) and without (dashed lines)
    grain growth considered. \label{fig:grain-growth}}
\end{figure}

Our analysis of the H$^{13}$CO$^{+}$~(1-0) emission observed in both
cores suggests that grain coagulation has occurred in the inner
regions of the cores. From the observational point of view, this is
consistent with the observations of core-shine in both objects
\citep{Pagani10}, that clearly indicates that \edit{larger grains
  (than average ISM grains)} are present. In B68, \citet{Bergin06b}
modeled the gas temperature structure and found that a reduced thermal
coupling between dust and grains -- probably owning to grain growth --
was needed in order to reproduce the gas temperature at the center of
the core. Additional evidence for grain growth in dense regions of
different clouds comes from the analysis of multi-wavelength FIR and
submillimeter observations \citep{Kramer03,Stepnik03,Juvella11}, and
is consistent with the expectation from theoretical models
\citep{Ossenkopf93,Ormel09}. Here we discuss the implications of grain
growth on the chemistry.

From the chemical point of view, grain growth reduces the average
grain cross section \edit{per H nucleus}, and therefore slows down
depletion. The timescale for depletion is given by:

\begin{equation}
  \tau_\mathrm{depletion}^{-1} = S \, n_\mathrm{grain} \,
  \sigma_\mathrm{grain} \, v_{th}
  \label{eq:5}
\end{equation}

\noindent where $S$ is the sticking coefficient of a molecule on the
grain, $n_\mathrm{grain}$ is the grain density, $\sigma_\mathrm{grain}$
is the grain cross section and $v_{th}$ is the gas thermal
velocity. For spherical grains with radius $a$, the grain density
writes as:

\begin{equation}
  n_\mathrm{grain} = \frac{3 \, m_\mathrm{H} \, m_\mathrm{dust/gas}}{4
    \, \pi \, a^{3} \, \rho_\mathrm{dust}} \, n_\mathrm{H}
  \label{eq:6}
\end{equation}

\noindent where $m_\mathrm{H}$ is the proton mass,
$m_\mathrm{dust/gas}$ is the dust to gas mass ratio, and
$\rho_\mathrm{dust}$ is the dust mass density. The grain cross
\edit{section} is simply:

\begin{equation}
  \sigma_\mathrm{grain} = \pi a^2
  \label{eq:7}
\end{equation}

\noindent
\edit{and the mean grain cross section per H nucleus is therefore:}

\begin{equation}
  \frac{n_\mathrm{grain} \, \sigma_\mathrm{grain}}{n_\mathrm{H}} =
  \frac{3 \, m_\mathrm{H} \, m_\mathrm{dust/gas}}{4 \, a \, \rho_\mathrm{dust}}
\end{equation}

\noindent From Eq.~(\ref{eq:5}), (\ref{eq:6}) and (\ref{eq:7}) we find
that the depletion timescale is proportional to the grains with
radius:

\begin{equation}
  \tau_\mathrm{depletion} \propto a
\end{equation}

In our \edit{modeling} we have assumed that $a$ varies as a power-law
of gas the density, above a certain density threshold. At the center
of L1498, we find a grain radius of $\editm{0.15}$~$\mu m$; therefore
the grain growth slows down depletion at the center of the core by
about \edit{50\%}.

\edit{Figure}~\ref{fig:grain-growth} compares the predicted abundances
of several species in L1498 for our best-fit model (including
grain-growth) and the same model with no grain-growth. The model with
grain-growth has a CO abundance at the center of the core which is
about a factor 4 larger than the model without grain-growth. For the
latter\edit{,} CO is depleted by about 3 orders of magnitude at the
core center with respect to its peak abundance. As a consequence, the
HCO$^{+}$ abundance at the center of the core is increased by about a
factor 4 when grain-growth is considered. On the other hand, we find
that it has little effect of the H$_3^+$ abundance: including grain
growth decreases slightly the H$_3^+$ abundance (which is partly
destroyed by CO). As a consequence, the partition of the main charge
carriers is not affected.

\section{Conclusions}
\label{sec:conclusions}

We presented observations of the C$^{18}$O~(1-0) and
H$^{13}$CO$^{+}$~(1-0) line emission in L1498 and L1517B. These
observations have been modeled with a detailed chemistry network
coupled with a radiative transfer code. Our observations and model
allow us to place constraints on the \edit{core} age \edit{and}
chemical composition. Our main conclusions are the following:

\begin{enumerate}

\item The C$^{18}$O~(1-0) emission is reproduced in both cores for a
  chemical core age of 0.3-0.5~Myr, similar in \edit{both sources}. We
  argue that these chemical ages are lower limits on time elapsed
  since the core has formed from their parent molecular
  cloud. Combining this lower limit with upper limits from
  \citet{Lee99} and \citet{Enoch08} core lifetimes, we \edit{estimate}
  that the core ages \edit{lie} within 0.3 and 1.6~Myr.  Our lower
  limits are comparable to the shortest lifetimes, suggesting that
  these cores are at a relatively evolved stage, and may shortly
  collapse to form protostars.

\item Our fiducial model underproduces the H$^{13}$CO$^{+}$~(1-0)
  emission by a about a factor 2. This line is found to probe deeper
  regions than C$^{18}$O~(1-0), because of both different excitation
  conditions (higher critical density) and chemistry. The discrepancy
  between the model and the observations is partly solved \edit{with}
  the inclusion \edit{of} cosmic-ray photodesorption. In addition, we
  find that an increase in the grain size in the inner regions of the
  cores is needed to reproduce the observations. This suggests that
  grain\edit{s} have coagulated in the innermost regions, in agreement
  with the observations of coreshine in both objects.

\item The predicted chemical structure of the cores can be roughly
  divided \edit{into} three regions: an external region where the
  chemistry is dominated by photo-dissociation and UV ionization; an
  intermediate region with undepleted abundance; and an internal
  region where the chemistry is dominated by freeze-out and cosmic-ray
  ionization. In the inner region, CO is depleted by about 2-3 orders
  of magnitude, and the ionization fraction is $0.7-1 \times 10^{-8}$.

\item Grain-growth is found to have an important effect on the
  chemistry; its main effect is to slow down the depletion by about
  \edit{50\%}. As a result\edit{,} the CO abundance in the inner
  regions is increased by a about a factor 4 in L1498. On the other
  hand, we found that it has no effect on the ionization fraction; in
  particular H$_3^{+}$ (and possibly its isotopologues) is always the
  main charge carrier.

\end{enumerate}

\begin{acknowledgements}
  The authors are grateful to Dimtri Semenov for his help for testing
  our chemical model. We also thank Robin Garrod for providing us the
  the binding energies of molecular species on dust grains, and Karin
  \"Oberg and Eric Herbst for discussions on
  photodesorption. \edit{Finally, we thank Malcolm Walmsley for his
    comments on this manuscript that helped to improve it.  This work
    was supported by NSF grant AST-1008800.}
\end{acknowledgements}

\bibliographystyle{aa}
\bibliography{bibliography}

\Online

\begin{appendix}

\section{Model predictions for the nitrogen chemistry}
\label{sec:model-pred-nitr}

\begin{figure*}
  \centering
  \begin{tabular}{cc}
    \includegraphics[width=8cm]{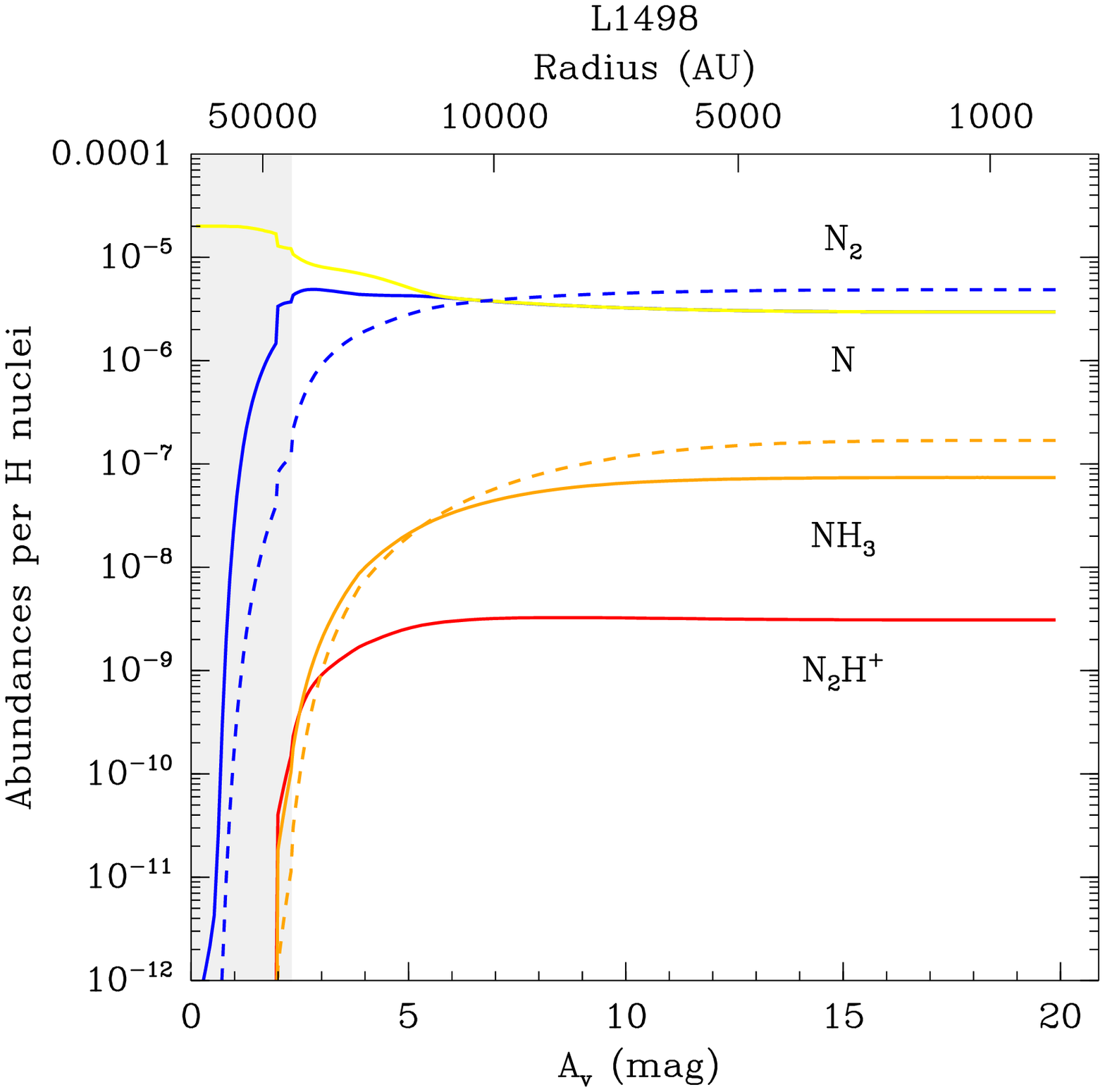} &
    \includegraphics[width=8cm]{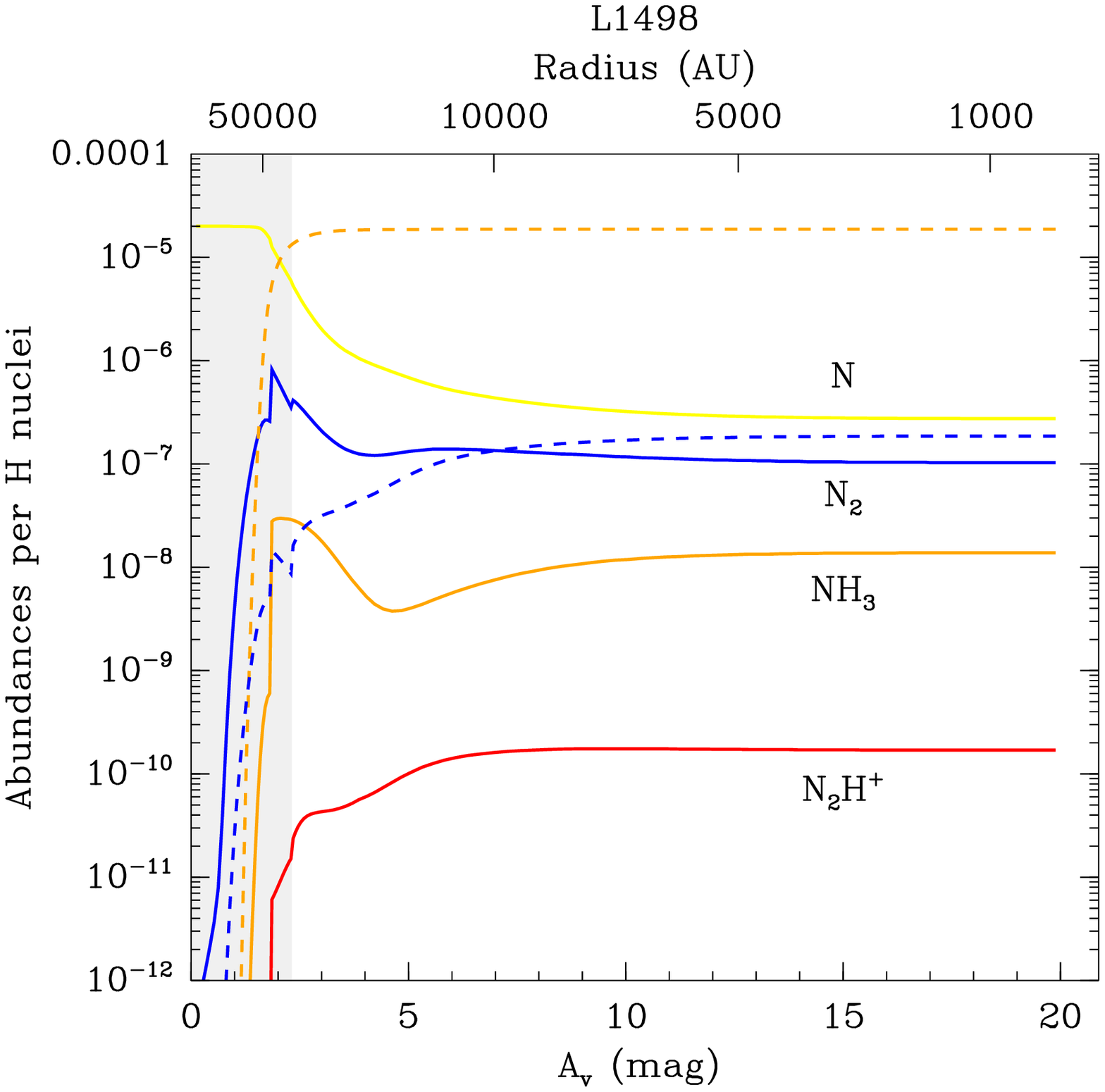} \\
  \end{tabular}
  \caption{\edit{Left panel: Predicted abundances for several nitrogen
      bearing species in L1498 for our best-fit model with an updated
      chemistry network (see text). The solid lines correspond to
      gas-phase abundances, and the dashed lines to ice
      abundances. Right panel: Same as in the left panel, but now
      assuming that most of nitrogen is initially in form of NH$_{3}$
      ices, with an abundance of 8.5\% relative to H$_{2}$O ices.}
  \label{fig:nitrogen}}
\end{figure*}

\edit{As discussed in Sect.~\ref{sec:cores-chem-struct}, our model
  best-fit model predicts that N$_{2}$H$^{+}$ contributes to about
  20-30\% of the total charge in both cores. Here we briefly discuss
  our model prediction for this species, and how it affects our
  results on the C$^{18}$O~(1-0) and H$^{13}$CO$^{+}$~(1-0) emission
  in the cores.  Our best fit model for L1498 predicts a
  N$_{2}$H$^{+}$ abundance of $3 \times 10^{-9}$, roughly constant
  throughout the core. This is more than an order of magnitude larger
  than the abundance derived by \citet{Tafalla04} in the same object
  ($9 \times 10^{-11}$ with respect to H nuclei), from
  N$_{2}$H$^{+}$~(1-0) and (3-2) line observations. This is likely
  because our model overproduces the N$_{2}$ abundance (a precursor of
  N$_{2}$H$^{+}$) which is the main nitrogen reservoir in our
  simulations. In order to reconcile our model predictions with the
  N$_{2}$H$^{+}$ observations, we have updated the rate of several
  reactions that are important for nitrogen bearing species. For the
  reaction rate of N$^{+}$ + H$_{2}$, we used the rate of
  \citet{Dislaire12}, assuming a H$_{2}$ ortho-to-para ratio of
  10$^{-3}$ \citep[the predicted steady-state value at 10~K;
  ][]{Faure13}. We have also updated the rates for the neutral-neutral
  reactions that lead to production of N$_{2}$ (see Table 3 from Le
  Gal et al., submitted to A\&A, and references therein). Finally, we
  have updated the branching ratio of the N$_{2}$H$^{+}$ dissociative
  recombination \citep{Vigren12}.}

\edit{The left panel of Fig.~\ref{fig:nitrogen} shows the predicted
  abundances of several nitrogen bearing species that we obtain for
  our best-fit model with a the updated chemistry network. We find
  that the new rates have no influence on the N$_{2}$H$^{+}$ abundance
  in the core center: although the N abundance is increased as a
  result of the reduced conversion of N to N$_{2}$ through neutral
  reactions, this has little influence on the N$_{2}$ abundance, which
  remains the main nitrogen reservoir.  To reduce the predicted
  N$_{2}$ abundance, we have run a model in which we assume that some
  of the nitrogen is initially in form of NH$_{3}$ ices, the rest of
  it being in atomic form. Assuming an initial NH$_{3}$ ice abundance
  of 5.5\% relative to H$_{2}$O ices \citep[the average value measured
  towards young stellar objects by ][]{Bottinelli10}, our model
  predicts a N$_{2}$H$^{+}$ abundance at the core center of $1 \times
  10^{-9}$, in better agreement with the observations, but still a
  factor 10 higher. Increasing the initial NH$_{3}$ ice abundance to
  8.5\% (still in agreement with the values measured by
  \citeauthor{Bottinelli10} that range between 2 and 15\%), our model
  predicts a N$_{2}$H$^{+}$ abundance of $2 \times 10^{-10}$, in
  reasonable agreement with the observations (see the right panel of
  Fig.~\ref{fig:nitrogen}). In addition, the predicted NH$_{3}$ gas
  phase abundance is $1 \times 10^{-8}$. This is also consistent with
  \citet[][]{Tafalla04}, who measured a para-NH$_3$ abundance of $7
  \times 10^{-9}$ (with respect to H nuclei) at the core center,
  i.e. a total (ortho-NH$_3$ + para-NH$_3$) abundance of $1 \times
  10^{-8}$, assuming an ortho-to-para ratio of 0.7 \citep{Faure13}. Of
  course, the model presented here should be compared to observations
  of other nitrogen bearing species, in order to obtain better
  constraints of the nitrogen chemistry as a whole (which is out of
  the scope if the present paper). However, it is important to note
  that it predicts the same C$^{18}$O~(1-0) and H$^{13}$CO~(1-0)
  integrated intensities as our best-fit model (within 10\%). Our
  conclusions on the CO and HCO$^{+}$ depletion are therefore
  independent of the assumptions made for the nitrogen chemistry.}

\end{appendix}

\end{document}